\documentclass[a4paper, 11pt]{article}
\pdfoutput=1 
\usepackage{graphicx}
\usepackage{jcappub} 
\usepackage{hyperref}
\usepackage{url}
\usepackage{slashed}
\usepackage{amsmath,amssymb} 
\usepackage{subcaption}
\allowdisplaybreaks

\begin{document}

\title{A Combined Astrophysical and Dark Matter Interpretation of the IceCube HESE and Throughgoing Muon Events }

\author[]{Yicong Sui,}
\author[]{P. S. Bhupal Dev}

\affiliation[]{Department of Physics and McDonnell Center for the Space Sciences, Washington University, St. Louis, MO 63130, USA}

\emailAdd{yicongsui@wustl.edu}
\emailAdd{bdev@wustl.edu}

\abstract{We perform a combined likelihood analysis for the IceCube 6-year high-energy starting events (HESE) above 60 TeV and 8-year throughgoing muon events above 10 TeV using a two-component neutrino flux model. The two-component flux can be motivated either from purely astrophysical sources or due to a beyond Standard Model contribution, such as decaying heavy dark matter. As for the astrophysical neutrinos, we consider two different source flavor compositions corresponding to the standard pion decay and muon-damped pion decay sources. We find that the latter is slightly preferred over the former as the high-energy component, while the low-energy component does not show any such preference. We also take into account the multi-messenger gamma-ray constraints and find that our two-component fit is compatible with these constraints, whereas the single-component power-law bestfit to the HESE data is ruled out. The astrophysical plus dark matter interpretation of the two-component flux is found to be mildly preferred by the current data and the gamma-ray constraints over the purely astrophysical explanation. }

\maketitle

\section{Introduction}\label{sec:intro}
The observation of ultra-high energy (UHE) neutrino events at the IceCube neutrino observatory~\cite{Aartsen:2013bka, Aartsen:2013jdh, Aartsen:2014gkd,  Aartsen:2015zva, Aartsen:2017mau} has commenced a new era in Neutrino Astrophysics~\cite{Anchordoqui:2013dnh}. In the 6 years of published data~\cite{Aartsen:2017mau}, IceCube has reported 82 high-energy starting events (HESE), i.e. events with the neutrino interaction vertex contained in the IceCube fiducial detector
volume. These include 22 track events generated in $\nu_\mu$-nucleon (and from a minority of $\nu_\tau$-nucleon) charged-current interactions, 58 shower events (either electromagnetic or hadronic) generated in both $\nu_e$ and $\nu_\tau$ charged-current as well as neutral-current interactions of any flavor, and 2 coincident events from unrelated cosmic ray (CR) air showers (that have been excluded from the analysis). Altogether, the 6-year HESE constitute $> 7\sigma$ excess over the expected atmospheric background of $25.2\pm 7.3$ muons and $15.6^{+11.4}_{-3.9}$ neutrinos, thus clearly pointing toward an extraterrestrial origin. Despite numerous searches by both IceCube~\cite{Aartsen:2016oji, Aartsen:2017kru} and ANTARES~\cite{Albert:2017uld}, the sources of these UHE neutrinos remain to be discovered. According to some recent studies~\cite{Denton:2017csz, Aartsen:2017ujz, Albert:2017oba, Pagliaroli:2017fse}, only a small fraction ($<14\%$ at 90\% CL) of these events could be attributed to a galactic origin;  therefore, most of the HESE must be extragalactic in nature. Moreover, no significant spatial clustering of the events was found~\cite{Aartsen:2017mau, Adrian-Martinez:2015ver} and the current data seems to be consistent with a diffuse isotropic neutrino flux from either uniformly distributed unresolved point sources or spatially extended  sources. There is also no significant correlation of the IceCube events with the arrival direction of UHECRs detected from the Pierre Auger or the Telescope Array~\cite{Aartsen:2015dml}. 

There are two conventional production mechanisms for diffuse astrophysical neutrinos from interactions of the CR protons and nuclei with the background gas or radiation present in a dense astrophysical environment, i.e. (i) hadro-nuclear production by inelastic $pp$ (or $np$) scattering in CR reservoirs like starburst galaxies and galaxy clusters/groups, where neutrinos are produced while they are confined within
the environment surrounding the CR source, and (ii) photo-hadronic production by $p\gamma$ scattering in CR accelerators like gamma-ray bursts, active galactic nuclei, radio galaxies, blazars, supernovae/hypernovae remnants and tidal disruption events, where neutrinos are produced within the CR source; see e.g. Refs.~\cite{Murase:2014tsa, Ahlers:2015lln, Meszaros:2017fcs} for recent reviews. Both $pp$ and $p\gamma$ sources produce a large amount of secondary charged pions and kaons, which subsequently undergo weak decays to produce neutrinos. Under this assumption, the neutrino+antineutrino flux at the source is expected to be distributed between the three neutrino flavors as $[(\nu_e+\bar{\nu}_e):(\nu_\mu+\bar{\nu}_\mu):(\nu_\tau+\bar{\nu}_\tau)]_{\rm S}=(1:2:0)$. Once emitted, they undergo vacuum oscillations over cosmic distances to produce equipartition in the three flavors (assuming tri-bi-maximal mixing) on their arrival at Earth: $[(\nu_e+\bar{\nu}_e):(\nu_\mu+\bar{\nu}_\mu):(\nu_\tau+\bar{\nu}_\tau)]_{\oplus}=(1:1:1)$~\cite{Learned:1994wg}. The   neutrino energies are related to the progenitor CR energies and the arrival direction of these neutrinos point back straight to their sources since neutrinos are neither deflected by magnetic fields unlike charged CRs nor absorbed by opaque matter unlike photons. Therefore, understanding the UHE neutrinos might provide new insights into the age-old problem of the origin and acceleration mechanism of UHECRs. 

To a first approximation, the astrophysical neutrino flux can be described by an unbroken isotropic power-law spectrum 
\begin{align}
\Phi(E_\nu) \ = \ \Phi_0 \left(\frac{E_\nu}{100~{\rm TeV}}\right)^{-\gamma} \, . 
\label{eq:1comp}
\end{align}
Using the $(1:1:1)_{\oplus}$ flavor composition and the deposited energy range of 60 TeV to 10 PeV, the 6-year HESE bestfit for the flux normalization (per flavor) and spectral index are respectively~\cite{Aartsen:2017mau} 
\begin{align}
\Phi_0 & \ = \ (2.46\pm 0.8)\times 10^{-18}~{\rm GeV}^{-1}{\rm cm}^{-2}{\rm s}^{-1}{\rm sr}^{-1}\, , \qquad \gamma  \ = \ 2.92_{-0.33}^{+0.29} \, . 
\label{eq:1compfit}
\end{align}

This fit, however, suffers from a number of issues:  
\begin{enumerate}
\item [(i)] {\it Fermi Shock Model:} The HESE spectrum is much softer than the theoretical prediction of $\gamma=2$~\cite{Waxman:1998yy}  from first-order Fermi diffusive shock acceleration mechanism~\cite{Fermi:1949ee, Fermi:1954ofk} (which may at most go up to $\gamma=2.5$~\cite{Spitkovsky:2008fi}); see Refs.~\cite{Blandford:1987pw, Baring:1997ka} for reviews.  In fact, the current bestfit $\gamma$ is softer compared to previous IceCube results~\cite{Aartsen:2013jdh, Aartsen:2014gkd,  Aartsen:2015zva}, because all of the HESE in the last two years have energies below 200 TeV. Moreover, lowering the energy threshold to 10 TeV hardens the spectral index to $2.50\pm 0.09$~\cite{Aartsen:2015knd}, which still disfavors the $E^{-2}$ spectrum at $3.8\sigma$. 

\item [(ii)] {\it Throughgoing Muons:} The HESE spectral index is also incompatible with the IceCube 8-year throughgoing (TG) muon sample~\cite{Aartsen:2017mau}. The TG muons come from muon neutrinos interacting outside the detector volume. In order to avoid the atmospheric background, the field of view is restricted to the Northern hemisphere (upward going events for IceCube) such that the atmospheric muons are absorbed in Earth. The sensitive energy range above which an extraterrestrial neutrino flux can be detected is about 200 TeV. The 8-year data sample contains almost 1000 extraterrestrial neutrinos above 10 TeV, which constitutes $6.7\sigma$ significance over the atmospheric-only hypothesis. When modeled as an isotropic power-law flux given by Eq.~\eqref{eq:1comp}, this yields the bestfit of~\cite{Aartsen:2017mau} 
\begin{align}
\Phi_{0,\mu} & \ = \ (1.01^{+0.26}_{-0.23})\times 10^{-18}~{\rm GeV}^{-1}{\rm cm}^{-2}{\rm s}^{-1}{\rm sr}^{-1} \, , \qquad \gamma \ = \ 2.19\pm 0.1 \, ,
\label{eq:1compfit2}
\end{align}
which suggests a harder spectrum close to the theoretically preferred $E^{-2}$. The spectral discrepancy between the HESE and TG datasets is approximately at $3\sigma$ level, cf. \eqref{eq:1compfit} and \eqref{eq:1compfit2}.

\item [(iii)] {\it Gamma-ray Constraint:} Both $pp$ and $p\gamma$ interactions lead to the production of charged and neutral pions. While the charged pions decay to give neutrinos, the neutral pions decay to give photons, which contribute to the high-energy diffuse gamma-ray spectrum in the TeV to PeV range and above\footnote{Very high-energy gamma-rays cascade against the infrared and microwave extragalactic background light (EBL), and could easily end up in the currently accessible energy range}. Since the gamma-ray and neutrino energies are related~\cite{Waxman:1997ti, Murase:2013rfa, Murase:2015xka}, the observed extraterrestrial neutrino flux leads to a calculable prediction of the gamma-ray flux; see Ref.~\cite{Becker:2007sv} for a review on the multi-messenger approach to UHE neutrinos. A comparison with the experimental upper limits on the all-sky diffuse gamma-ray flux seems to disfavor the single-component bestfit given by Eq.~\eqref{eq:1compfit}. See Section~\ref{sec:gamma} for more details.  
\end{enumerate}

These issues have led us to question the single power-law hypothesis~\eqref{eq:1comp} and instead entertain the possibility of a break in the neutrino spectrum~\cite{Chen:2014gxa, Anchordoqui:2016ewn, Vincent:2016nut, Palladino:2016zoe, Palladino:2016xsy, Chianese:2017jfa, Palladino:2017qda, Palladino:2018evm}, analogous to (and may be corresponding to~\cite{Candia:2003ay, Gandhi:2005at, Murase:2008yt, Guo:2013rya, Palladino:2018evm}) the break in the CR spectrum~\cite{DeRujula:2018kfg}. A two-component flux can arise either from (a) purely astrophysical sources, such as one galactic and one extragalactic component~\cite{Marinelli:2016mjo, Neronov:2015osa, Palladino:2016zoe, Denton:2017csz, Palladino:2018evm}, or (b) due to a some beyond Standard Model (SM)  contribution to the astrophysical neutrino flux, e.g. from heavy dark matter (DM) decay~\cite{Esmaili:2013gha, Bai:2013nga, Bhattacharya:2014vwa, Rott:2014kfa, Esmaili:2014rma, Aisati:2015vma, Anchordoqui:2015lqa, Brdar:2016thq, Kuznetsov:2016fjt, Bhattacharya:2016tma, Hiroshima:2017hmy, Bhattacharya:2017jaw, Chianese:2016opp, Chianese:2016kpu, Chianese:2017nwe, Ema:2013nda, Aartsen:2018mxl}. For specific model examples of decaying heavy DM, see e.g. Refs.~\cite{Feldstein:2013kka, Higaki:2014dwa, Fong:2014bsa, Boucenna:2015tra, Ko:2015nma, Fiorentin:2016avj, Borah:2017xgm, Dev:2016qbd, Dhuria:2017ihq, DiBari:2016guw, Chianese:2016smc, Chakravarty:2017hcy}. The decaying heavy DM interpretation was initially invoked to explain the mild excess in the PeV energy range and the lack of events in the multi-PeV range, especially near the Glashow resonance of 6.3 PeV~\cite{Glashow:1960zz, Bhattacharya:2011qu, Barger:2012mz}. 
However, after the publication of the TG data sample with one event at $(2.6\pm 0.3)$ PeV deposited energy which corresponds to $8.7$ PeV median expected muon neutrino energy~\cite{Aartsen:2016xlq}, the PeV DM hypothesis seems less favored. On the other hand, the attention has now shifted to the 40-200 TeV region of the HESE data sample, which exhibits a $\sim 2\sigma$ excess, with respect to a single power-law with $\gamma=2$~\cite{Chianese:2016opp, Chianese:2016kpu, Chianese:2017nwe}.\footnote{It is interesting to note that the ANTARES 9-year shower data~\cite{Albert:2017bdv} also exhibits a slight excess in the same energy range, although the overall significance of their events with respect to the expected atmospheric background is below $2\sigma$. } This becomes an important feature if one wants to simultaneously explain the HESE and TG data samples, because the TG data prefers a hard spectrum close to $E^{-2}$, cf. Eq.~\eqref{eq:1compfit2}.  

In this paper, we make a comparative study of the two-component hypothesis for the purely astrophysical and astrophysical+DM scenarios. For the flavor composition ratio of the astrophysical neutrinos at the source, we consider the canonical $(1:2:0)$, as well as the muon-damped $(0:1:0)$ case. For the DM-induced component, we assume a simplified fermionic DM model with a direct Yukawa coupling to the neutrinos. A statistical likelihood analysis is performed with the combined IceCube 6-year HESE and 8-year TG data sets~\cite{Aartsen:2017mau}. We find that 
\begin{enumerate}
\item [(i)] For the purely astrophysical two-component flux, the current data prefers a low-energy component of the form $e^{-E/E_c}$ with a cut-off around $E_c\sim 100$ TeV and a $\sim E^{-2}$ high-energy component without any cut-off. 

\item [(ii)] For the astrophysical+DM case, the single-unbroken power-law flux in Eq.~\eqref{eq:1comp} with $\gamma=2$ is consistent with the data, if augmented by a low-energy component from DM decay. For the example DM model we have considered, the bestfit values of the DM mass and lifetime turn out to be $M_{\rm DM}\:=\:315^{+335}_{-125} \:\, $TeV and $\tau_{\rm DM}\:=\:6.3^{+12.7}_{-2.3}\times 10^{28}\rm\, sec$, respectively, which are consistent with existing cosmological constraints on decaying DM. 

\item [(iii)] As for the astrophysical neutrino flavor composition at source, the goodness of fit for the $(0:1:0)$ case is slightly better than the $(1:2:0)$ case for both astrophysical and astrophysical+DM scenarios. This is mainly due to the absence of shower events in the Glashow bin and the presence of  two  TG muon events in the multi-PeV range, which favors a larger muon-neutrino flux compared to the electron (or tau)-neutrino flux. 

\item [(iv)] Between the purely astrophysical and astrophysical+DM cases, the goodness of fit is slightly better for the latter case. This is mainly due to the $2\sigma$ excess in the 100 TeV bin of HESE data.  

\item [(v)] Our bestfit prediction of the integrated gamma-ray flux for the astrophysical+DM case is consistent with the current experimental upper bound in both $pp$ and $p\gamma$  scenarios. The corresponding bestfit prediction for the purely astrophysical two-component case is also consistent with the gamma-ray flux limit, but on the verge of being excluded. On the other hand, the bestfit single-component astrophysical case is clearly excluded by the gamma-ray constraint. 
\end{enumerate}

The rest of the paper is organized as follows: in Section~\ref{sec:astro}, we analyze a purely astrophysical two-component flux; in Section~\ref{sec:astro-dm}, we assume one of the components as coming from DM decay; in Section~\ref{sec:gamma}, we discuss the multi-messenger constraints from gamma-rays. Our conclusions are given in Section~\ref{sec:conc}. 


\section{Two-component Astrophysical Spectrum}~\label{sec:astro}
The possibility of having multiple components for the astrophysical neutrino flux, rather than having a single component as in Eq.~\eqref{eq:1comp}, seems quite plausible, given the fact that despite a large number of searches for the origin of the IceCube events, no point sources have been identified so far. Galactic sources  cannot contribute more than a few \% to the total flux~\cite{Denton:2017csz, Aartsen:2017ujz, Albert:2017oba, Pagliaroli:2017fse}. Even for the extragalactic contribution, although there exist several candidates~\cite{Murase:2014tsa, Ahlers:2015lln, Meszaros:2017fcs}, none of them seems to be able to explain {\it all} of the IceCube events, while being consistent with the multi-messenger constraints from gamma-rays. In particular, prompt emission from triggered gamma-ray bursts are strongly constrained to $<1\%$ of the total flux~\cite{Aartsen:2016oji}. Blazars are constrained to contribute less than 27\% of the flux for $E^{-2.5}$ (or 50\% for $E^{-2.2}$)~\cite{Aartsen:2016lir}. Similarly, the contribution of star-forming galaxies can be at most 22\% in order to satisfy the Fermi-LAT extragalactic background~\cite{Bechtol:2015uqb, Sudoh:2018ana}. Therefore, it is natural to expect that a combination of more than one kind of sources, and hence, a multi-component flux, might be responsible for the IceCube events. 

Here we will consider a simple two-component flux~\cite{Chen:2014gxa}: 
\begin{equation}
\Phi(E_{\nu}) \ = \ \Phi_{1_0} \left(\frac{E_\nu}{100~{\rm TeV}}\right)^{-\gamma_1}e^{-E_\nu/E_c} \: + \: \Phi_{2_0}\left(\frac{E_\nu}{100~{\rm TeV}}\right)^{-\gamma_2} ,
\label{eq:2comp}
\end{equation} 
where the second component is similar to the single-component power-law in Eq.~\eqref{eq:1comp}, while the first component has an exponential cut-off energy scale $E_c$. The spectral break ($\gamma_1\neq \gamma_2$) could arise, for instance, in some astrophysical models due to a steepening of the CR diffusion coefficient~\cite{He:2013cqa, Murase:2013rfa, Liu:2013wia, Senno:2015tra}. Similarly, the exponential cut-off could be understood as due to the presence of a spectral resonance, as e.g. in $p\gamma\to \Delta^+$, or due to an intrinsic dissipative source cut-off such as gamma-ray bursts~\cite{Murase:2013ffa, Petropoulou:2014lja, Asano:2015kma}. The exact value of the energy cut-off in the neutrino spectrum is unknown, although correlating it to the corresponding knee of the CR spectrum at about 3 PeV for protons, one would obtain $E_c\simeq 150$ TeV, assuming that the proton component dominates the neutrino production and considering that the average neutrino energy is roughly $5\%$ of the primary proton energy~\cite{Kelner:2006tc}. In our model-independent approach, we will treat $E_c$ as a free parameter, along with the spectral indices $\gamma_1,\gamma_2$ and the flux normalizations $\Phi_{1_0}, \Phi_{2_0}$, and derive their bestfit values from a statistical likelihood analysis of the combined HESE and TG data samples from IceCube. 

\subsection{Flavor Composition} \label{sec:flavor}

As for the flavor composition ratio of the astrophysical neutrinos, for $pp$ interactions, 
\begin{align}
 pp \ & \to \ X\pi^\pm \, , \qquad 
\pi^+ \ \to \ \mu^+\nu_\mu \ \to \ e^+\nu_e \bar{\nu}_\mu \nu_\mu \, , \quad 
 \pi^- \ \to \ \mu^-\bar{\nu}_\mu \ \to \ e^-\bar{\nu}_e\nu_\mu \bar{\nu}_\mu \, ,
\label{eq:decay1}
\end{align}
(where $X$ stands for other hadrons) leading to the source flavor composition of 
\begin{align}
(\nu_e: \nu_\mu: \nu_\tau: \bar{\nu}_e: \bar{\nu}_\mu:\bar{\nu}_\tau)_{\rm S} \ = \ \left(\frac{1}{6}:\frac{1}{3}:0:\frac{1}{6}:\frac{1}{3}:0\right) \, .
\label{eq:ratio1}
\end{align}
For $p\gamma$ interactions, we could have either the direct production $p\gamma\to X\pi^\pm$ leading to the same flavor ratio as Eq.~\eqref{eq:ratio1} or the resonant production
\begin{align}
p\gamma \ & \to \ \Delta^+ \ \to \ \left\{\begin{array}{c} p\pi^0 \ \to \ p\gamma\gamma ~~~~~~~~~~~~~~~~~~~~~~~~ \\
 n\pi^+  \ \to \ n\mu^+\nu_\mu \ \to \ ne^+\nu_e\bar{\nu}_\mu \nu_\mu 
\end{array}\right. \, ,
\label{eq:decay2}
\end{align}
leading to the source flavor composition of 
\begin{align}
(\nu_e: \nu_\mu: \nu_\tau: \bar{\nu}_e: \bar{\nu}_\mu:\bar{\nu}_\tau)_{\rm S} \ = \ \left(\frac{1}{3}:\frac{1}{3}:0:0:\frac{1}{3}:0\right) \, .
\label{eq:ratio2}
\end{align}
The main difference here is that at the $\Delta^+$-resonance, $p\gamma$ interactions  produce only $\pi^+$ and no $\pi^-$, which results in only $\nu_e$ and no $\bar{\nu}_e$ production. This is one way to explain the absence (or suppression\footnote{A small amount of $\pi^-$ can be produced either from the multi-pion process $p\gamma\to n\pi^-+X\pi^+\pi^-$, or from the back-reaction process $n\gamma\to p\pi^-$, depending on the optical depth of the source. The presence of free neutrons could also give rise to additional $\bar{\nu}_e$'s from the $\beta$-decay process $n\to pe^-\bar{\nu}_e$, either inside or outside the source environment depending on the free neutron energy. For such decay, however, the $\bar{\nu}_e$ carries on average 0.1\% of the parent neutron energy, so to produce the Glashow resonance, the original proton energy must be in the EeV range corresponding to a very small CR flux.}) of the Glashow resonance $\bar{\nu}_ee^-\to W^-$ at $E_\nu\simeq m_W^2/2m_e\simeq 6.3$ PeV~\cite{Glashow:1960zz}, if this feature in the current data persists with more statistics.

Note that in both $pp$ and $p\gamma$ cases, the electron (anti)neutrinos are produced from muon decays. Since the rest-frame lifetime of muons, $\tau_\mu=2.2\times 10^{-6}$ sec, is about 85 times larger than that of charged pions, $\tau_{\pi^\pm}=2.6\times 10^{-8}$ sec, it is possible that the muons in the decay chain of Eq.~\eqref{eq:decay1} or Eq.~\eqref{eq:decay2} rapidly lose energy in the source environment, e.g. due to synchrotron radiation in a strong magnetic field or by multiple scattering in the dense astrophysical medium, before decaying~\cite{Rachen:1998fd, Kashti:2005qa, Winter:2012xq}. In this muon-damped source environment, the source flavor compositions in Eqs.~\eqref{eq:ratio1} and \eqref{eq:ratio2} respectively become
\begin{align}
(\nu_e: \nu_\mu: \nu_\tau: \bar{\nu}_e: \bar{\nu}_\mu:\bar{\nu}_\tau)_{\rm S}\ = \ \left\{\begin{array}{cc} \left(0:\frac{1}{2}:0:0:\frac{1}{2}:0\right) ~~~~~~~~~~~~~  & (pp) \\
(0:1:0:0:0:0) ~~~~~~~~~~~~~ & (p\gamma) 
\end{array}\right. 
\, .
\label{eq:ratio3}
\end{align} 
Note that in this case also, no electron antineutrinos are produced (for both $pp$ and $p\gamma$ sources), and this is another way to explain the absence of events in the Ghashow bin.

After escaping the source environment, the individual neutrino flavors undergo vacuum oscillations over cosmic distances before reaching the detector on Earth. Given a flavor ratio $f_{\ell, {\rm S}}$ of the neutrino species $\nu_\ell$ at source, the corresponding value $f_{\ell,\oplus}$ on Earth is given by 
\begin{align}
f_{\ell, \oplus} \ = \ \sum_{\ell'=e,\mu,\tau} \sum_{i=1}^3 |U_{\ell i}|^2|U_{\ell' i}|^2 f_{\ell', {\rm S}} \ \equiv \ \sum_{\ell'} P_{\ell \ell'} f_{\ell',{\rm S}} \, ,
\label{eq:pmns}
\end{align}
where $U_{\ell i}$ are the elements of the PMNS mixing matrix and $P_{\ell \ell'}$ is the vacuum oscillation probability for $\nu_\ell \to \nu_{\ell'}$. As the IceCube detector cannot distinguish neutrinos from antineutrinos (except for the Glashow resonance or when the matter effects are important), we simply sum over the per flavor neutrino and antineutrino fluxes at the detector. Using Eq.~\eqref{eq:pmns} and taking the current bestfit values of the 3-neutrino oscillation parameters~\cite{Esteban:2016qun, nufit}, we get the corresponding flavor composition ratios on Earth: 
\begin{align}
(\nu_e+\bar{\nu}_e):(\nu_\mu+\bar{\nu}_\mu):(\nu_\tau+\bar{\nu}_\tau) \ = \ \left\{\begin{array}{cc} (1:1:1)_{\oplus} & {\rm for}~(1:2:0)_{\rm S} \\
(4:7:7)_{\oplus} & {\rm for}~(0:1:0)_{\rm S} 
\end{array}\right. \, ,
\label{eq:flavor}
\end{align}
which is valid for both  $pp$ and $p\gamma$ sources. 
In what follows, we will consider these two possibilities, both of which are consistent with the current IceCube data at $3\sigma$~\cite{Aartsen:2017mau}.\footnote{Another possibility, namely, a neutron-rich source giving rise to $(1:0:0)$ flux at source which becomes $(5:2:2)$ on Earth is now ruled out at $>3\sigma$ CL~\cite{Aartsen:2017mau}.} So we modify Eq.~\eqref{eq:2comp} to include the possibility of having different flavor compositions for the two components: 
\begin{equation}
\Phi_{\nu_\ell}(E_\nu) \ = \ f_{1,\ell}\Phi_{1_0} \left(\frac{E_\nu}{100~{\rm TeV}}\right)^{-\gamma_1}e^{-E_\nu/E_c}\:+\: f_{2,\ell}\Phi_{2_0}\left(\frac{E_\nu}{100~{\rm TeV}}\right)^{-\gamma_2} \, ,
\label{eq:2compf}
\end{equation}   
where $f_{1,\ell}$ and $f_{2,\ell}$ are the flavor composition factors on Earth given by Eq.~\eqref{eq:pmns} for a known flavor factor $f_{\ell,{\rm S}}$ at the source. 

\subsection{Likelihood Analysis}\label{sec:reconstruct} 

We use the two-component flux given by Eq.~\eqref{eq:2comp} with five free parameters, namely, the cut-off energy $E_c$, the spectral indices $\gamma_{1,2}$ and the flux normalizations $\Phi_{1_0,2_0}$, to calculate both HESE and TG event spectra at IceCube. We then perform a statistical likelihood analysis with the combined 6-year HESE and 8-year TG data samples to find the bestfit values of these five parameters. 

Given the astrophysical neutrino flux~\eqref{eq:2comp}, the total number of expected HESE in each deposited energy bin can be calculated using  
\begin{equation}\label{eq:reconst2comp}
N^{\rm HESE}_i \ = \ \int d\Omega \int_{E_{i,{\rm min}}}^{E_{i, {\rm max}}}\:dE \ \sum_{\ell=e,\mu,\tau} \Phi_{\nu_\ell}(E) \cdot T \cdot A_{\nu_\ell}(E,\Omega) \, ,
\end{equation}
where $\Omega$ is the solid angle of coverage, $T$ is the exposure time (equal to 2078 days for the 6-year HESE data), and $A_{\nu_\ell}$ is the effective area per energy per solid angle for the neutrino flavor $\nu_\ell$~\cite{Aartsen:2013jdh, Aartsen:2017mau}, assuming SM neutrino-nucleon scattering cross section. $E$ is the electromagnetic-equivalent deposited energy, which is always smaller than the incoming neutrino energy $E_\nu$ in the laboratory frame by a factor depending on $E_\nu$ and the type of interaction~\cite{Kistler:2013my, Chen:2013dza, Aartsen:2013vja, Palomares-Ruiz:2015mka, Palladino:2018evm}. Based on the numerical data provided in Ref.~\cite{Palladino:2018evm}, we obtain a linear relation between the deposited energy $E$ and the average value of actual energy $E_\nu$ (both in GeV units) valid in the range of 10 TeV - 10 PeV deposited energy: 
\begin{align}
E_{\nu} & \ = \ \left\{\begin{array}{cc} 34.4~{\rm GeV}\:+\:1.105\:E & (\text{shower})\\
-104.4~{\rm GeV}\:+\:3.745\:E & (\text{track}) 
\end{array}\right. .
\label{eq:reconstruct}
\end{align}
In Eq.~\eqref{eq:reconst2comp}, the limits of the energy integration $E_{i,{\rm min}}$ and $E_{i,{\rm max}}$ give the size of the $i$th deposited energy bin over which the expected number of events is being calculated. Note that for an isotropic flux (as assumed here), the flux $\Phi_{\nu_\ell}$ given by Eq.~\eqref{eq:2compf} does not depend on the solid angle. 

For the TG events, the corresponding expected number of events in the $i$th energy bin is given by 
\begin{equation}\label{eq:reconst2comptg}
N^{\rm TG}_i \ = \ \int d\Omega \int_{E_{i,{\rm min}}}^{E_{i, {\rm max}}}\:dE \ \sum_{\ell=e,\mu,\tau} \Phi_{\nu_\ell}(E) \cdot F_{\nu_\ell}(E,\Omega) \, ,
\end{equation}
which is similar to Eq.~\eqref{eq:reconst2comp}, but with the $T\cdot A_{\nu_\ell}(E,\Omega)$ factor replaced by the exposure function $F_{\nu_\ell}(E,\Omega)$ as given for the 8-year TG data sample~\cite{Aartsen:2017mau}. On the other hand, the energy reconstruction from the deposited energy to the incoming neutrino energy for the TG events is highly nontrivial and has a large uncertainty~\cite{Aartsen:2013vja, Palladino:2018evm}; so we just use the median value of the reconstructed energy as given by the IceCube collaboration~\cite{Aartsen:2017mau} in Eq~\eqref{eq:reconst2comptg}. Similar to the $\nu_e$ case, we assume a linear relation between the median energy and the real energy, which however varies from bin to bin and cannot be expressed in a simple universal form like Eq.~\eqref{eq:reconstruct}. 


After we obtain the expected signal events $N_{{\rm astro},i}$ from astrophysical neutrinos using Eq.~\eqref{eq:reconst2comp} for HESE and Eq.~\eqref{eq:reconst2comptg} for TG events, we add the corresponding background expectations $N_{{\rm atm},i}$ as given by the IceCube collaboration~\cite{Aartsen:2017mau} to obtain the total number of events in the $i$th energy bin, i.e. 
\begin{equation}
N_{{\rm tot},i} \ = \ N_{{\rm astro},i}+N_{{\rm atm},i} \, .
\label{eq:tot}
\end{equation}
To compare the reconstructed events bin by bin with the observed HESE and TG data, we do a combined goodness of fit test with the parameter set $\theta=\{\Phi_{1_0},\Phi_{2_0},\gamma_1,\gamma_2,E_c\}$.
Our test statistic (TS) is chosen as 
\begin{equation}
\begin{split}
\text{TS}\ = \  -2\ \text{ln}[\lambda(\theta)_{\rm HESE}\cdot\lambda(\theta)_{\rm TG}]\ =\ & 2 \sum_{i=1}^{i_{\rm max}} \left[{N^{\rm HESE}_{{\rm tot},i}} (\theta)-n^{\rm HESE}_i + n^{\rm HESE}_i \ \text{ln}\left(\frac{n^{\rm HESE}_i}{{N^{\rm HESE}_{{\rm tot},i}} (\theta)}\right)\right] \\ & + 2\sum_{j=1}^{j_{\rm max}} \left[{N^{\rm TG}_{{\rm tot},j}} (\theta)-n^{\rm TG}_j + n^{\rm TG}_j\  \text{ln}\left(\frac{n^{\rm TG}_j}{{N^{\rm TG}_{{\rm tot},j}} (\theta)}\right)\right] 
\end{split}
\label{eq:TS}
\end{equation}
where $n_{i}^{\rm HESE}$ and $n_{j}^{\rm TG}$ are the number of HESE and TG events as reported by IceCube in the ${i}$th and $j$th energy bin, respectively. We assume that ${N^{\rm HESE}_{{\rm tot},i}}$ and ${N^{\rm TG}_{{\rm tot},i}}$ each follows a Poissonian distribution, and hence, the TS follows a $\chi^2$-distribution, which is then used to infer the $2\sigma$ and $3\sigma$-preferred regions of the $\theta$-parameter space. Note that the binning for HESE and TG events is done differently by IceCube and we have treated the datasets accordingly. 

\subsection{Fit Results\label{2comprecon}}
As mentioned in Section~\ref{sec:flavor}, we consider two possibilities for the flavor compositions of the astrophysical neutrino flux, cf.~Eq.~\eqref{eq:flavor}. Thus, for our two-component flux, there are four distinct combinations, where each component can have either $(1:1:1)$ or $(4:7:7)$ on Earth. In what follows, we show the fit results for two cases, namely, with the low-energy (1st) component being $(1:1:1)$ while the high-energy (2nd) component being either $(1:1:1)$ or $(4:7:7)$. As we will show below, the 1st component makes a sub-dominant contribution to the overall fit; therefore, the remaining two cases with the 1st component being $(4:7:7)$ yield practically the same results as those shown here.

\begin{figure}[t!]
	\centering
	\begin{subfigure}[b]{0.40\textwidth}
		\centering
		\includegraphics[width=\textwidth]{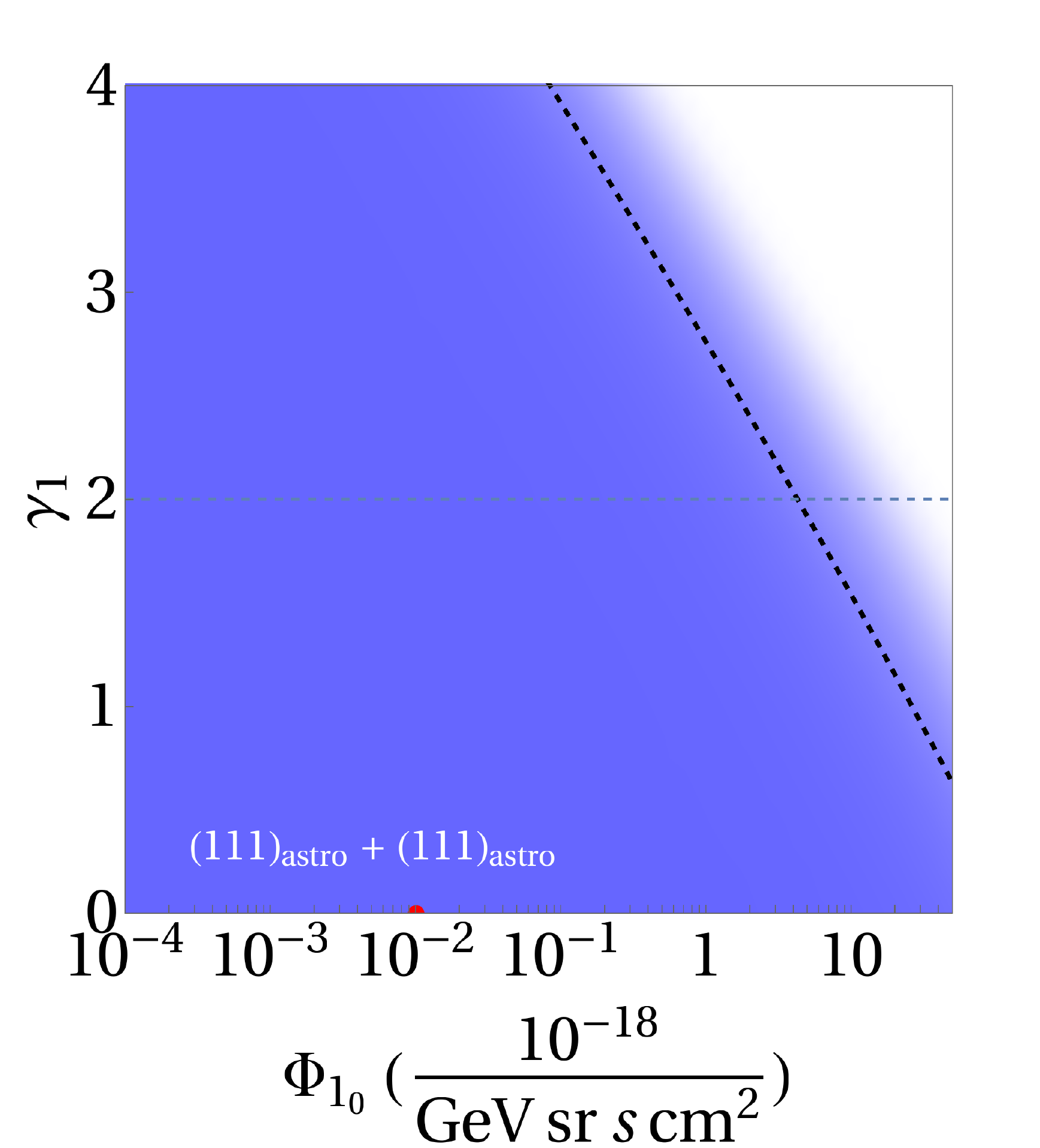}
		\caption[Network2]%
		{{\small $(\Phi_{1_{0}},\gamma_1)$ map for (1:1:1) + (1:1:1). }}    
	\end{subfigure}
	\quad
\begin{subfigure}[b]{0.40\textwidth}   
		\centering 
		\includegraphics[width=\textwidth]{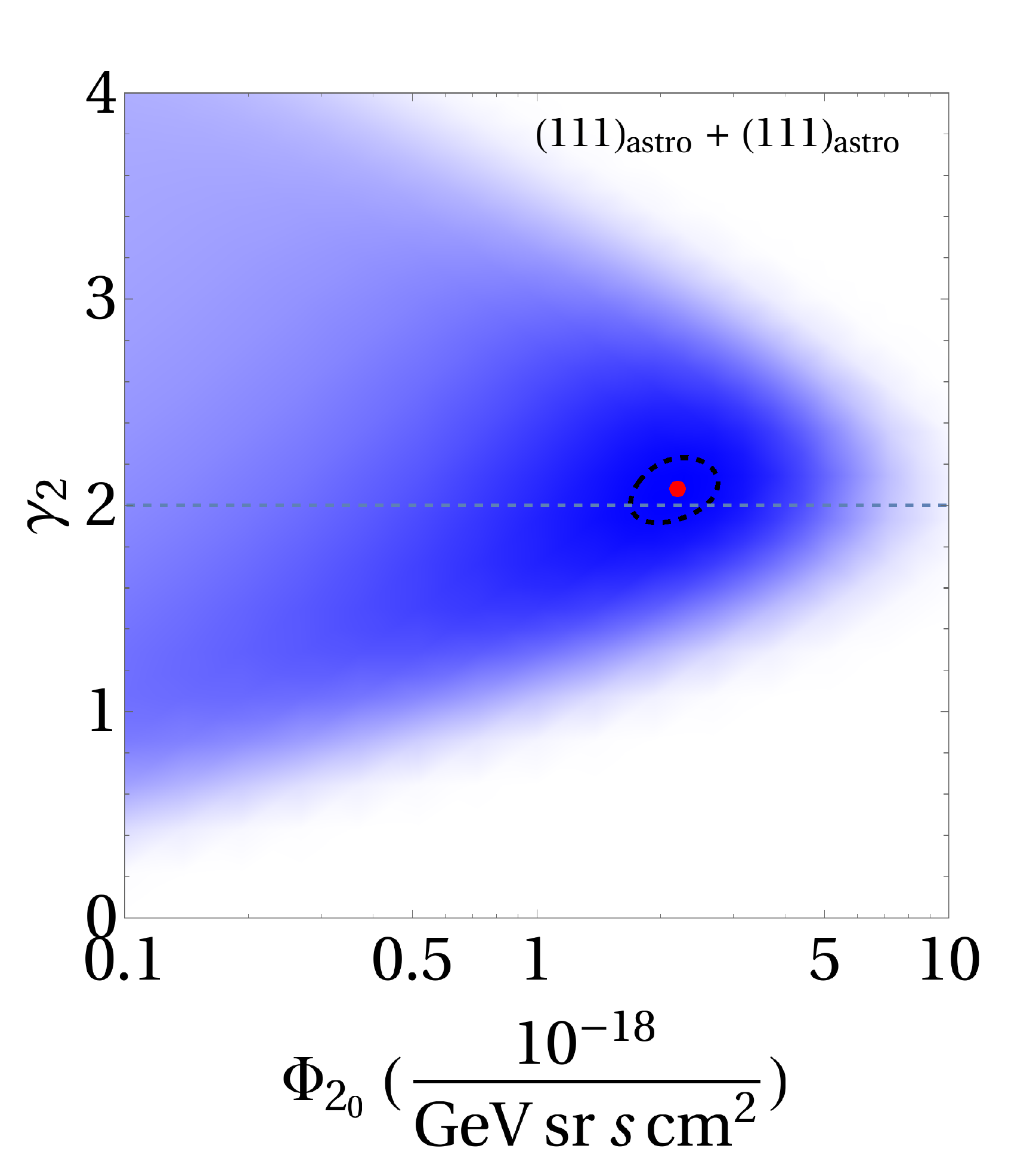}
		\caption[]%
		{{\small $(\Phi_{2_{0}},\gamma_2)$ map for (1:1:1) + (1:1:1).  }}    
	\end{subfigure}
\vskip\baselineskip
	\begin{subfigure}[b]{0.40\textwidth}  
		\centering 
		\includegraphics[width=\textwidth]{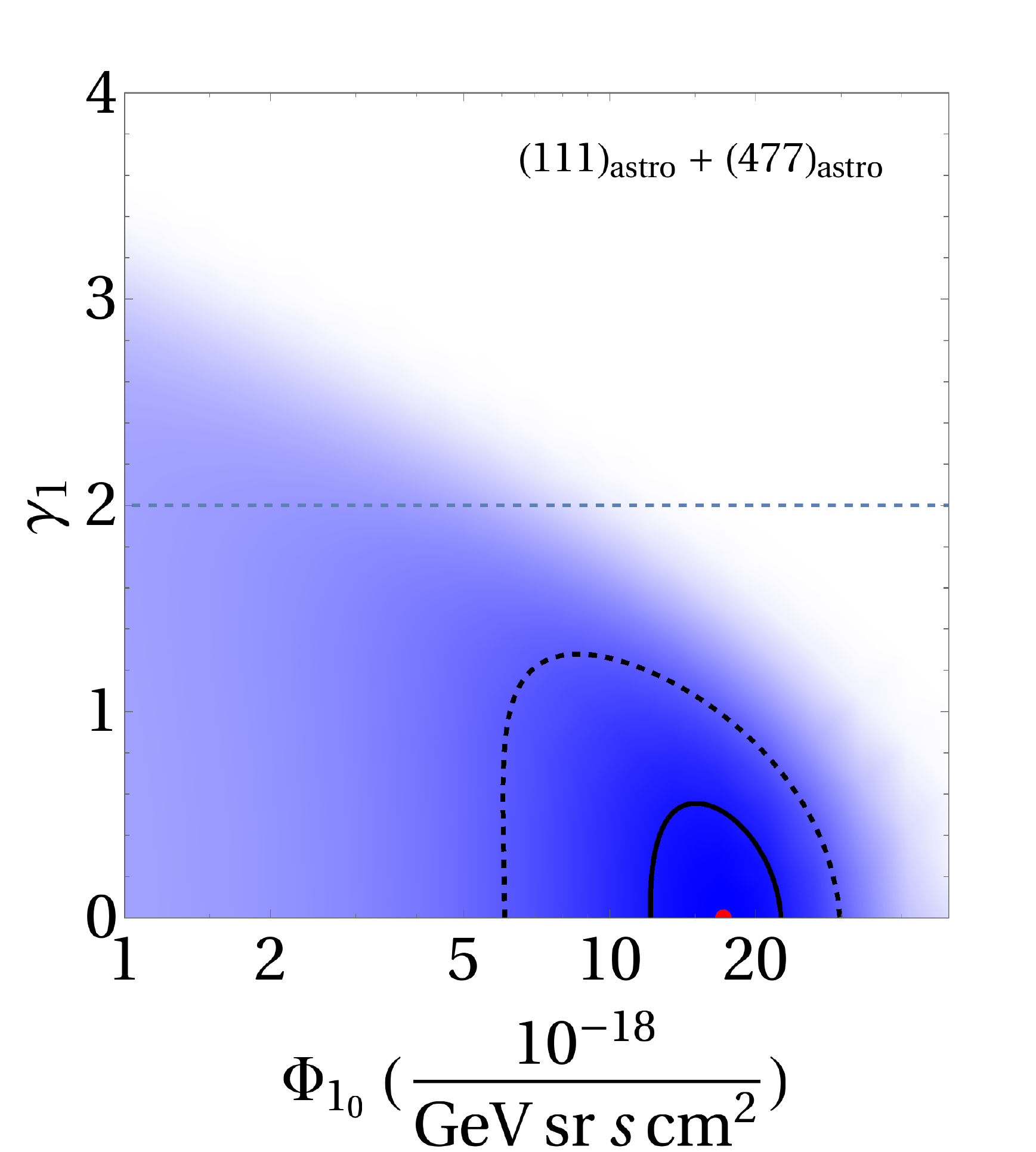}
		\caption[]%
		{{\small $(\Phi_{1_{0}},\gamma_1)$ map for (1:1:1) + (4:7:7).  }}    
	\end{subfigure}
	\quad
	\begin{subfigure}[b]{0.40\textwidth}   
		\centering 
		\includegraphics[width=\textwidth]{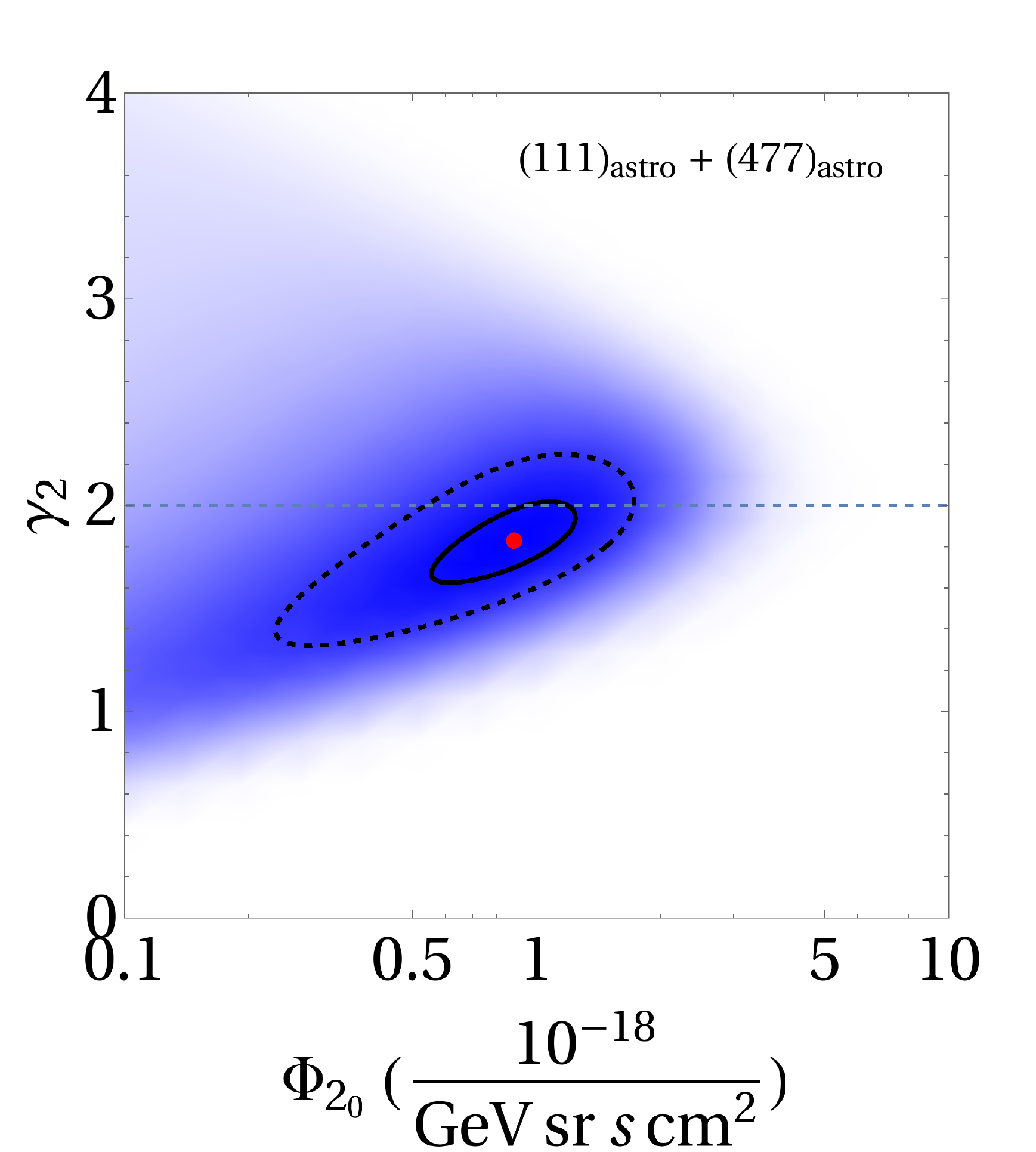}
		\caption[]%
		{{\small $(\Phi_{2_{0}},\gamma_1)$ map for (1:1:1) + (4:7:7).}}    
	\end{subfigure}
	\caption[]
	{\small Preferred regions of parameter space in the $(\Phi_{1_{0}},\gamma_1)$ [left panels] and $(\Phi_{2_{0}},\gamma_1)$ [right panels] planes for the two-component astrophysical neutrino flux, cf. Eq.~\eqref{eq:2compf}, with $(1:1:1) + (1:1:1)$ [top panels] and $(1:1:1) + (4:7:7)$ [bottom panels] flavor compositions. The solid (in the bottom panels) and dashed contours show the $2\sigma$ and 3$\sigma$ preferred ranges respectively, while the red dot in each map shows the bestfit value as given in Table~\ref{tab:1}.} 
	\label{fig:2compTSmap}
\end{figure}

We perform the joint likelihood analysis following the procedure outlined in Section~\ref{sec:reconstruct} and construct the test statistics using Eq.~\eqref{eq:TS}. The bestfit values of the five parameters $\{\Phi_{1_0},\Phi_{2_0},\gamma_1,\gamma_2,E_c\}$, along with the corresponding TS per degree of freedom (dof) are shown in Table~\ref{tab:1} for the two cases mentioned above. In Figure~\ref{fig:2compTSmap}, we show the allowed ranges in the $(\Phi_{1_0},\gamma_1)$ and $(\Phi_{2_0},\gamma_2)$ planes for both these cases. Here we have fixed the value of $E_c$ at the bestfit value given in Table~\ref{tab:1} for each case. The $3\sigma$-preferred ranges are shown by the dashed contours, while in Figure~\ref{fig:2compTSmap} (c) and (d), the $2\sigma$ range can also be seen by the solid contours. The theoretically preferred value of $\gamma=2$ is shown by the horizontal dashed line.

\begin{table}[h!]
	\centering
	\begin{tabular}{l l| c c c c c |c } 
		\hline\hline
		1st Comp. & 2nd Comp. & $\Phi_{1_{0}}$ & $\Phi_{2_{0}}$ & $\gamma_1$ & $\gamma_2$ & $E_c/100~{\rm TeV}$ & TS/dof \\ [0.5ex] 
		\hline
	 $(1:1:1)$ & $(1:1:1)$ & 0.01 & 2.21 & 1.47$\times10^{-4}$ & 2.08 & 0.10 & 1.91 \\ 
		
	 $(1:1:1)$ & $(4:7:7)$ & 17.18 & 0.88 & 3.19$\times10^{-10}$ & 1.83 & 0.50 & 1.48 \\[1ex]
		\hline\hline
	\end{tabular}
	\caption{Bestfit results for the two-component astrophysical neutrino flux, cf.~Eq.~\eqref{eq:2compf}. Here $\Phi_{1_0}$ and $\Phi_{2_0}$ are in units of $10^{-18}/({\rm GeV\: sr\: s\: cm}^2)$. \label{tab:1}}
	
\end{table}

For the $(1:1:1)+(1:1:1)$ case, we see from Figure~\ref{fig:2compTSmap} (a) that the bestfit values of both the parameters of the first component $(\phi_{1_0},\gamma_1)$ are closer to zero, which means the first component is basically not contributing to the events at all. Therefore, this fit actually becomes a  `one-component fit' with $\Phi_0\simeq 2.2\times10^{-18}({\rm GeV\: sr\: s\: cm}^2)^{-1}$ and $\gamma \simeq 2.1$, as shown in the first row of Table~\ref{tab:1}. Note that this is slightly different from the two-component fit given by the IceCube collaboration~\cite{Aartsen:2017mau}, because they fix the high-energy component to be the TG bestfit as a prior, whereas we treat both components as free parameters in our fit. In any case, our bestfit value of $\gamma$ is closer to the TG-alone bestfit given by Eq.~\eqref{eq:1compfit2}, and not the HESE-alone bestfit given by Eq.~\eqref{eq:1compfit}, simply because the TG data sample is much larger than the HESE sample, and hence, the statistics is dominated by the TG data. In other words, we expect this fit to better explain the observed TG event spectrum, but not the HESE spectrum. 

\begin{figure}[t!]
	\centering
	\begin{subfigure}[b]{0.48\textwidth}
		\centering
		\includegraphics[width=\textwidth]{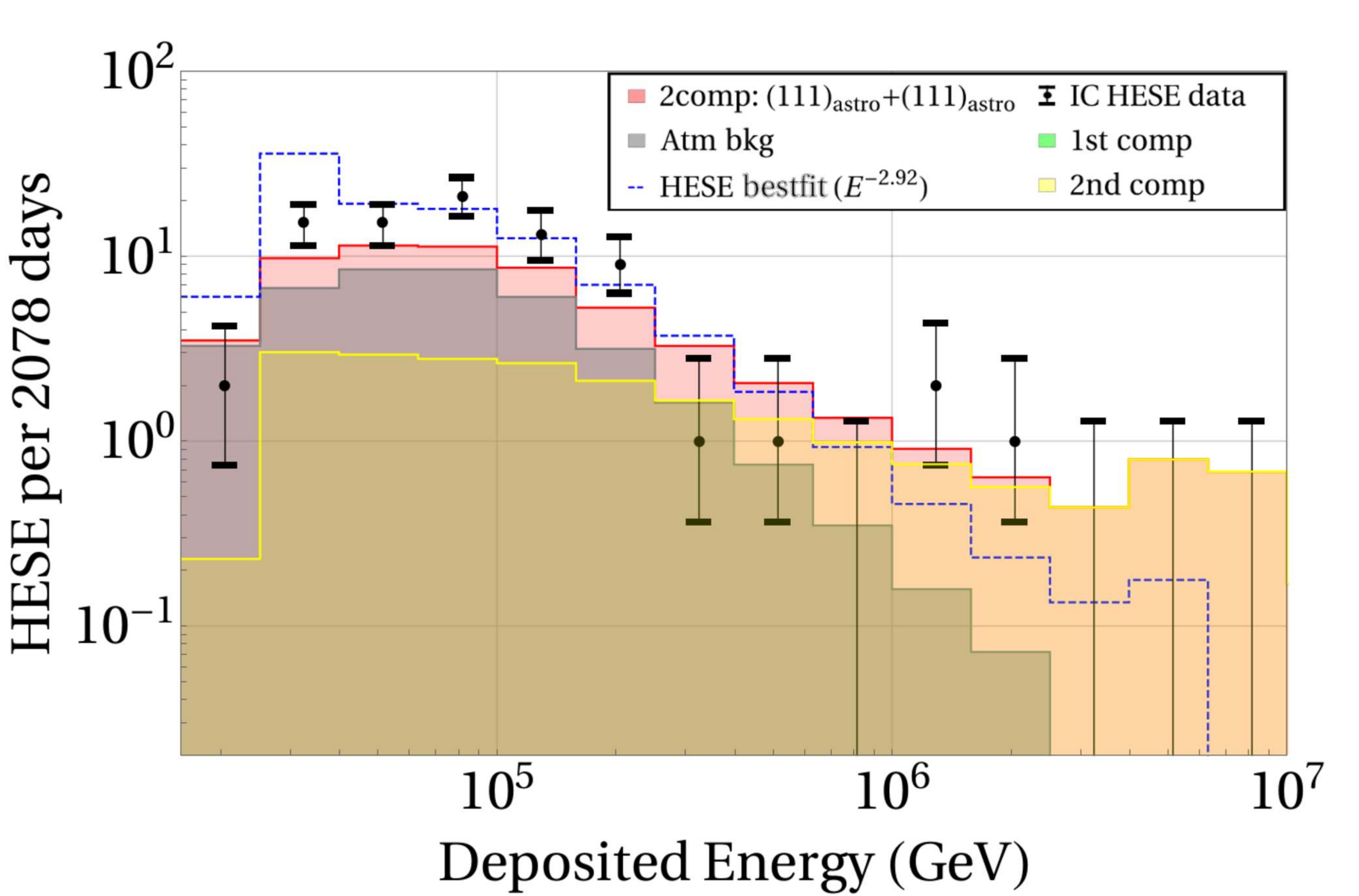}
		\caption[]%
		{{\small Combined bestfit predictions for HESE with $(1:1:1) + (1:1:1)$ and comparison with the one-component fit.}}    
		\label{fig:1111112compEventHESE}
	\end{subfigure}
	\quad
	\begin{subfigure}[b]{0.48\textwidth}  
		\centering 
		\includegraphics[width=\textwidth]{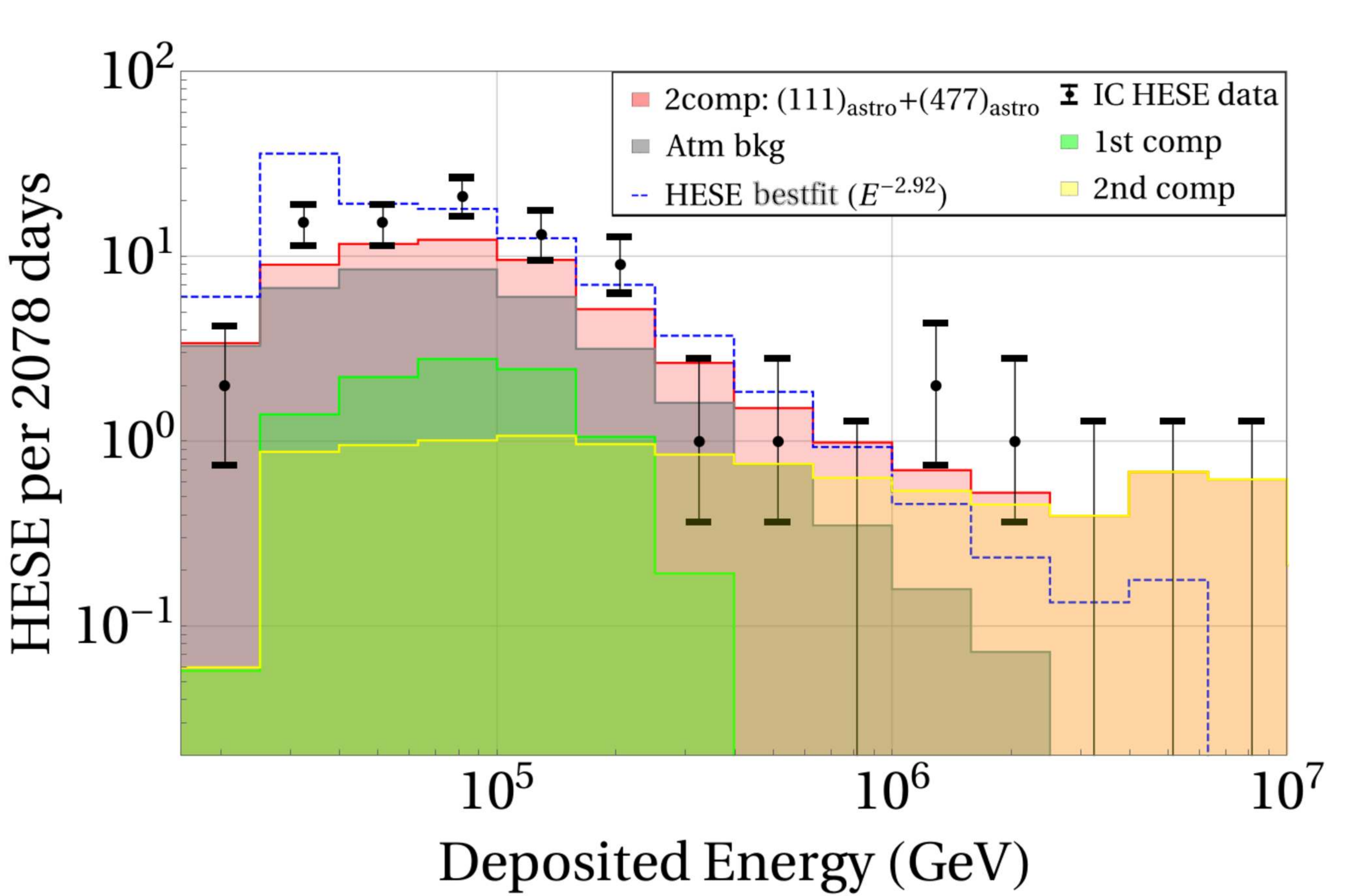}
		\caption[]%
		{{\small Combined bestfit predictions for HESE with $(1:1:1) + (4:7:7)$ and comparison with the one-component fit.}}    
		\label{fig:1114772compEventHESE}
	\end{subfigure}
	\vskip\baselineskip
	\begin{subfigure}[b]{0.48\textwidth}   
		\centering 
		\includegraphics[width=\textwidth]{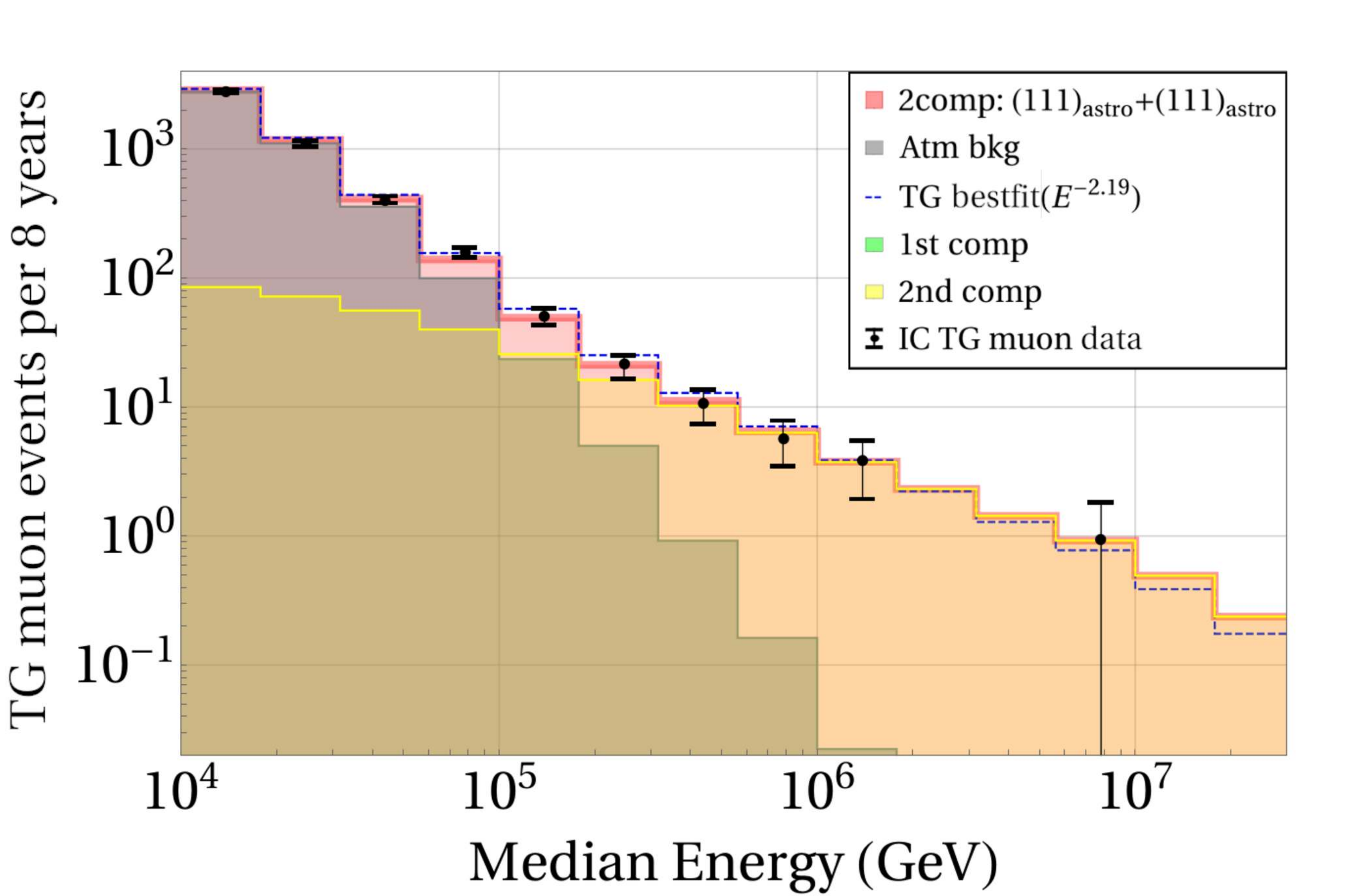}
		\caption[]%
		{{\small Combined bestfit predictions for TG muon events with $(1:1:1) + (1:1:1)$ and comparison with the one-component fit.}}    
		\label{fig:1111112compEventTG}
	\end{subfigure}
	\quad
\begin{subfigure}[b]{0.48\textwidth}   
		\centering 
		\includegraphics[width=\textwidth]{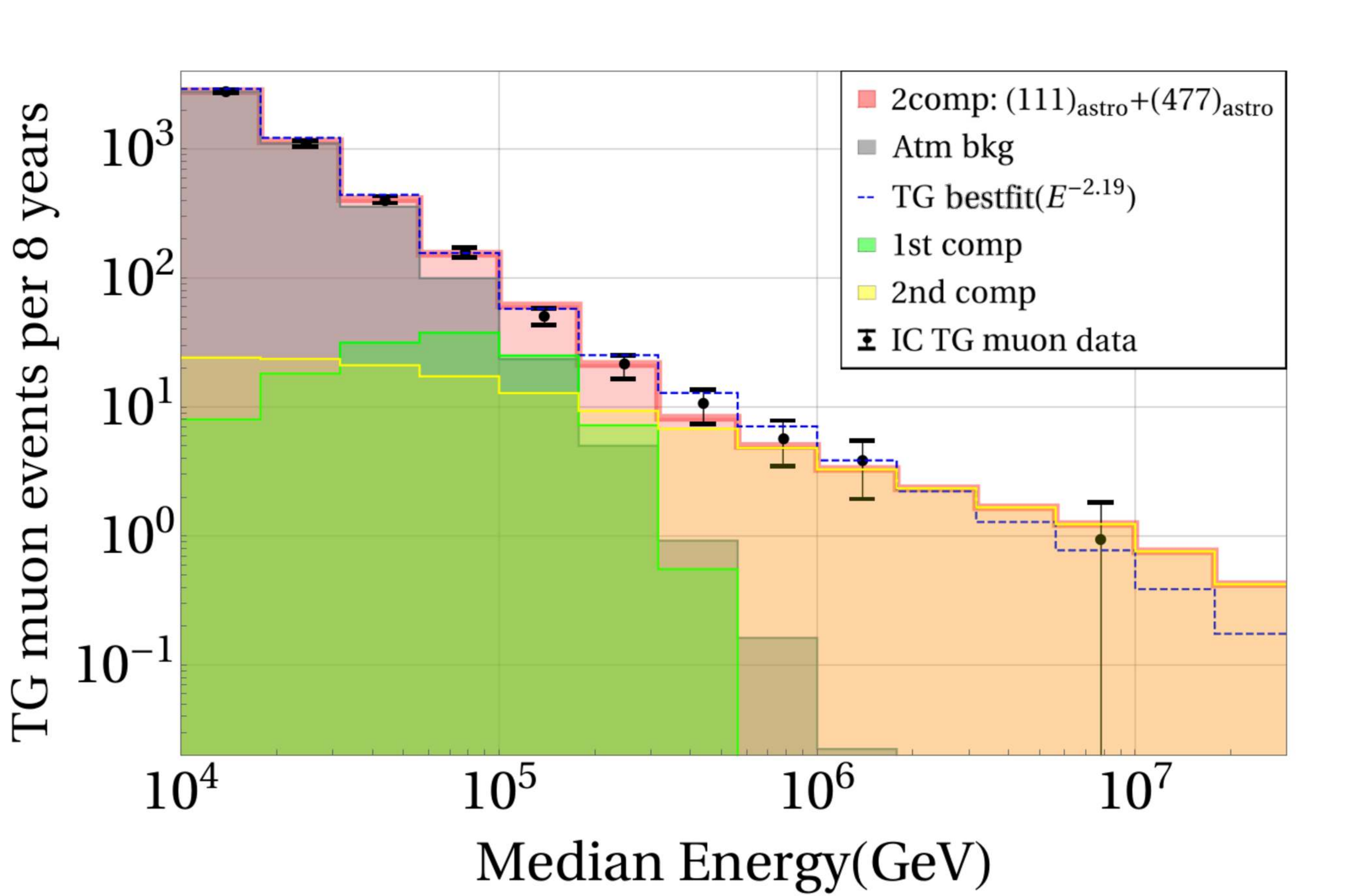}
		\caption[]%
		{{\small Combined bestfit predictions for TG muon events with $(1:1:1) + (4:7:7)$ and comparison with the one-component fit.}}    
		\label{fig:1114772compEventTG}
	\end{subfigure}
	\caption[]
	{\small Event spectrum for the two-component bestfit reconstruction of HESE and TG data samples. The grey shaded part is the contribution due to atmospheric background and the pink shaded part is the total contribution (two-component astro + bkg), with the green and yellow shades the individual contributions from first and second components respectively. In the $(1:1:1)+(1:1:1)$ case (left panels), the first component is vanishingly small [cf. Table~\ref{tab:1}], so there is no green shaded region. The dashed blue curve in (a) and (b) is the one-component HESE-alone bestfit [cf.~Eq.~\eqref{eq:1compfit}], while the dashed blue curve in (c) and (d)  is the one-component TG-alone bestfit [cf.~Eq.~\eqref{eq:1compfit2}]. The IceCube data points for 6-year HESE and 8-year TG samples are also shown.} 
	\label{fig:2compEvents}
\end{figure}

This is explicitly shown in Figure~\ref{fig:2compEvents} (a) and (c), where we show the combined reconstructed two-component event spectra with $(1:1:1)+(1:1:1)$ for the HESE and TG data samples, respectively, using the bestfit values given in Table~\ref{tab:1}.  Here the grey shaded part is the contribution due to atmospheric background, as reported by IceCube~\cite{Aartsen:2017mau} and the pink shaded part is the total contribution (two-component astro + bkg), which is essentially the contribution from the second component (yellow shaded) plus the atmospheric background. Note that here the first component is vanishingly small [cf.~Table~\ref{tab:1}], so there is no green shaded region. The dashed blue curves in Figure~\ref{fig:2compEvents} (a) and (c) are for the one-component HESE-alone bestfit [cf.~Eq.~\eqref{eq:1compfit}] and TG-alone bestfit [cf.~Eq.~\eqref{eq:1compfit2}], respectively. The IceCube data points for 6-year HESE and 8-year TG samples are also shown. It is clear that our two-component flux with $(1:1:1)+(1:1:1)$ is a good fit to the TG data (essentially same as the one-component fit), but not to the HESE data, especially in the low-energy bins. This is due to the harder spectral index in our case, compared to Eq.~\eqref{eq:1compfit}. 

For the $(1:1:1)+(4:7:7)$ case, we see from Figure~\ref{fig:2compTSmap} (c) that for the first component, the  bestfit spectral-index is still close to zero, while the flux normalization is no longer small, which means the first component does contribute to the event reconstruction at lower energies and exponentially dies off at higher energies. This can be easily seen from Figure~\ref{fig:2compEvents} (b) and (d) where the green shaded region represents the contribution from the first component. The corresponding bestfit parameter values are given in Table~\ref{tab:1} last row. Note that here the spectral index for the second component is even harder, cf. Figure~\ref{fig:2compTSmap} (d), but still it provides a better overall fit to both HESE and TG data, as can be seen by the lower value of TS/dof, as compared to the $(1:1:1)+(1:1:1)$ case. This is because a harder flux provides a better fit to the high-energy bins, while the non-negligible contribution from the first component, along with the second component, provides a better fit to the low-energy bins. Thus, we conclude from this section that the $(1:1:1)+(4:7:7)$ case is slightly preferred over the $(1:1:1)+(1:1:1)$ case for the purely astrophysical two-component flux, and we hope that with more statistics, one could clearly distinguish these from each other. 

As for the reason why in the $(1:1:1)+(1:1:1)$ case, the data forces upon us the one-component model, even though we started with a two-component model, whereas in the $(1:1:1)+(4:7:7)$ case, it remains as a two-component model, this has to do with the detailed spectral features in Figure~\ref{fig:2compEvents}. First of all, a $pp$ $(1:1:1)$ high-energy component would contribute to the Glashow bin, whereas a $(4:7:7)$ component does not, because it does not have an electron antineutrino component [cf. Eqs.~\eqref{eq:ratio2} and \eqref{eq:ratio3}]. Thus the absence of Glashow events in the data forces the second component to have a softer spectrum for the $(1:1:1)$ case, while the $(4:7:7)$ case could still afford a harder spectrum. This is clear from Figure~\ref{fig:2compTSmap} (b) and (d). Secondly, since we let the cut-off energy scale as a free parameter, it is purely determined by the test statistics. Since a softer spectrum predicts more events in the low-energy bins, there is no need for a separate low-energy component, and therefore, the data prefers the first component being negligible in the $(1:1:1)$ case. On the other hand, for a harder spectrum, as in the $(4:7:7)$ case, one component is not enough to explain all the low-energy events, and that's why, the data prefers a non-negligible low-energy component. In the future data with more statistics, if the paucity of events in the Glashow bin continues, it might be a strong indication of either a $(4:7:7)$ flavor composition (either $pp$ or $p\gamma$) or a $p\gamma$ source for the high-energy component. 


In any case, for both the two-component fits shown in Figure~\ref{fig:2compEvents} (a) and (b), there is a visible $\sim 2\sigma$ excess in the current HESE data. This appears as a generic feature if we attempt to explain both HESE and TG data simultaneously. A possible particle physics explanation of this excess is given in the following section.

\section{A Decaying Dark Matter Component}\label{sec:astro-dm}
There is overwhelming astrophysical and cosmological evidence for the existence of DM, which constitutes 27\% of the energy budget (or 85\% total mass) of our universe~\cite{Patrignani:2016xqp}. But the nature  and  properties  of  DM  are still unknown, which are among the most important open questions in physics. If the DM thermalizes with the SM particles in the early universe, there is a well-known partial wave unitarity upper bound of ${\cal O}(20)$ TeV on its mass~\cite{Griest:1989wd, Nussinov:2014qva}. However, there exist ways to push the unitarity limit to higher DM masses, see e.g. Refs.~\cite{Dev:2016xcp, Berlin:2016gtr, Dev:2016qeb, Bramante:2017obj, Baldes:2017gzw}. Moreover, the DM need not be absolutely stable and all we require is that it must be stable on time-scales much longer than the age of the universe. In this context, it is interesting to ask whether a heavy decaying DM could contribute to the IceCube neutrino events. This idea has been entertained for the HESE data with a PeV-scale DM  in Refs~\cite{Esmaili:2013gha, Bai:2013nga, Bhattacharya:2014vwa, Rott:2014kfa, Esmaili:2014rma, Aisati:2015vma, Anchordoqui:2015lqa, Brdar:2016thq, Kuznetsov:2016fjt, Bhattacharya:2016tma, Hiroshima:2017hmy, Bhattacharya:2017jaw, Chianese:2016opp, Chianese:2016kpu, Chianese:2017nwe, Ema:2013nda, Aartsen:2018mxl}, and with a 100 TeV scale DM in Refs.~\cite{Chianese:2016opp, Chianese:2016kpu, Chianese:2017nwe}. In light of the $\sim 2\sigma$ excess in the HESE data around 100 TeV [cf. Figure~\ref{fig:2compEvents} (a) and (b)] when we attempt to simultaneously explain the HESE and TG events, we will consider in this section the possibility of a few hundred TeV scale DM decaying directly into neutrinos. Because of the long decay lifetime, and hence, extremely small couplings of the DM to the SM particles, we expect such DM particles to be non-thermally produced in the early universe, e.g. from the decay of inflaton~\cite{Allahverdi:2002nb, Giudice:2004ce, Dev:2013yza, Harigaya:2014waa, Drees:2017iod}, or through the freeze-in mechanism~\cite{Hall:2009bx} to give the correct relic density.

\subsection{A Simple Model\label{test}}
For concreteness, we adopt a simple model of heavy fermionic DM ($\chi$) which directly decays to the SM neutrinos, so that we could get potentially large contributions to the IceCube events. Since the SM neutrinos are part of the $SU(2)_L$ doublet, the only possible renormalizable interaction is of the form 
\begin{align}
-{\cal L}_{\rm Y} \ = \ y_{i}\bar{L}_i\tilde{\phi} \chi+ {\rm H.c} \, ,
\label{eq:yuk}
\end{align}
where $\phi$ and $L$ are respectively the SM Higgs and lepton doublets, and $\tilde{\phi}=i\sigma^2\phi^*$, with $\sigma^2$ being the second Pauli matrix. Together with a Majorana mass term $M_{\rm DM}\chi^TC^{-1}\chi$, where $C$ is the charge conjugation operator, the DM field resembles a right-handed neutrino in the seesaw model~\cite{Minkowski:1977sc, Mohapatra:1979ia, Yanagida:1979as, GellMann:1980vs, Glashow:1979nm}. However, the cosmologically long lifetime of the DM requires the Yukawa couplings to be extremely small: $y_i\sim {\cal O}(10^{-30})$ for PeV-scale DM~\cite{Higaki:2014dwa, Esmaili:2014rma, Ko:2015nma, DiBari:2016guw}; so it does not contribute significantly to the neutrino masses.

\begin{figure}[t!]	
	\centering	
    \includegraphics[width=\textwidth]{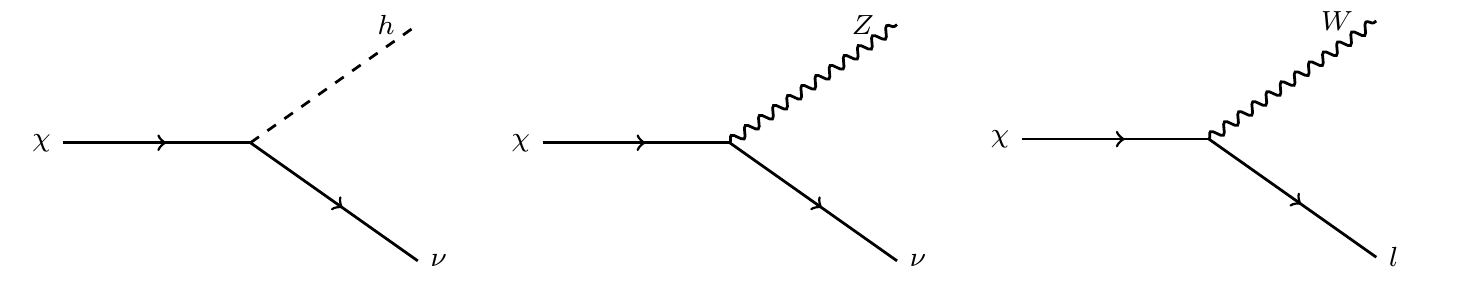}
	\caption{DM decays to SM particles induced by the interaction Lagrangian~\eqref{eq:Lint}. Since we assume $\chi$ to be a Majorana particle, it also decays to the corresponding antiparticles with equal probability. }
\label{fig:feyn}
\end{figure}

After electroweak symmetry breaking, the Yukawa Lagrangian~\eqref{eq:yuk} induces the following couplings to the SM $W$, $Z$ and Higgs bosons~\cite{Atre:2009rg}: 
\begin{equation}
 -\mathcal{L}_{\rm int} \ = \ V_{\ell \chi}\left(\frac{g}{\sqrt{2}}\:W^+_\mu\:\bar{\chi}\:\gamma^{\mu}P_{L}\ell^{-} + \frac{g}{2\:{\rm cos}\:\theta_{w}}\:Z_\mu\:\bar{\chi}\:\gamma^{\mu}P_{L}\nu_{\ell} + \frac{g\cdot { M_{\rm DM}}}{2\:{ M_W}}\:\textit{h}\:\bar{\chi}P_{L}\nu_{\ell}\right)+ {\rm H.c.} \, ,
\label{eq:Lint}
\end{equation}
where $V_{\ell \chi}\simeq yv/M_{\rm DM}$ is the $\chi-\nu$ mixing term in the seesaw approximation, $v$ is the electroweak vacuum expectation value, $g$ is the $SU(2)_L$ gauge coupling and $\theta_w$ is the weak mixing angle. The interaction Lagrangian~\eqref{eq:Lint} 
 induces the two-body decays $\chi\to h\nu_\ell, Z\nu_\ell, W\ell$, as shown in Figure~\ref{fig:feyn}. 
For simplicity, we set all the $V_{\ell\chi}$'s equal for different lepton flavors, i.e. the DM decays to all neutrino flavors with the same branching ratio, and hence, the flavor composition of the neutrinos+antineutrinos at source is $(1:1:1)$, which after oscillation also remains $(1:1:1)$ on Earth [cf.~Eq.~\eqref{eq:pmns}]. 

As we can see from Figure~\ref{fig:feyn}, high energy neutrinos are directly produced from the DM decay to $Z\nu$ and $h\nu$. Assuming $ M_{\rm DM}\gg M_{W}$, we would approximately have $E_{\nu}\sim{ M_{\rm DM}}/{2}$ for these primary neutrinos. Thus, to explain the $E\sim 100$ TeV peak in Figure~\ref{fig:2compEvents} (a) and (b), we need $ M_{\rm DM}\sim 200$ TeV. As we will see later, the bestfit value of $M_{\rm DM}$ is within a factor of few of this naive estimate. Note that some secondary neutrinos will also be produced from the $W$ and $Z$ decays, but these will be of lesser energy than the primary ones. We include both primary and secondary neutrinos in our numerical analysis.



In the model described above, the DM mass $M_{\rm DM}$ and the effective coupling $V_{\ell \chi}$ are the only new free parameters introduced. We could trade off the parameter $V_{\ell \chi}$ in favor of the total lifetime $\tau_{\rm DM}=1/\Gamma_{\rm DM}$, where $\Gamma_{\rm DM}$ is the total decay width (in the limit $M_{\rm DM}\gg M_{W}$): 
\begin{equation}
\begin{split}
\Gamma_{\rm DM} \ \simeq \ \frac{3g^2}{16\pi}|V_{\ell \chi}|^2  \frac{M^3_{\rm DM}}{M^2_{W}} \, .
\end{split}
\end{equation}
So our fit results will be shown in the $(M_{\rm DM},\tau_{\rm DM})$ plane. 


\subsection{Neutrino Flux from DM Decay}
Assuming that we have the $(1:1:1)$ neutrino flux $\Phi_{\rm DM}$ from DM decay as described above, together with a single-component unbroken astrophysical power-law flux $\Phi_{\rm astro}$ given by Eq.~\eqref{eq:1comp}, whose flavor composition could be either $(1:1:1)$ or $(4:7:7)$ on Earth, we obtain an effective two-component neutrino flux: 
\begin{equation}
\Phi_{\rm tot} \ = \ \Phi_{\rm DM}+ \Phi_{\rm astro} \, .
\label{eq:dmastro}
\end{equation}
As for the DM flux, we include both galactic and extragalactic contributions:
\begin{equation}
\Phi_{\rm DM} \ = \ \Phi_{\rm G}+ \Phi_{\rm EG} \, .
\end{equation} 
The Galactic contribution, averaged over all directions, is given by
\begin{equation}
\Phi_{\rm G} \ = \ \frac{1}{4\pi M_{\rm DM} \tau_{\rm DM}}\int \frac{dN(E_\nu)}{dE_{\nu}}\ \frac{d\Omega(l,b)}{4\pi} \int_{0}^{\infty} ds\ \rho_{\rm DM}[r(s,l,b)] \, ,
\end{equation}
where $dN/dE_\nu$ is the neutrino energy spectrum from DM decay, $\Omega(l,b)$ is the solid angle, $(s,l,b)$ are the galactic coordinates ($l$ is the longitude, $b$ is the latitude and $s$ is the distance to the Sun), $\rho_{\rm DM}$ is the DM density profile in Milky Way, for which we assume a Navarro-Frenk-White (NFW) profile~\cite{Navarro:1996gj}: 
\begin{align}
\rho_{\rm DM}(r) \ = \ \frac{\rho_0}{(r/r_s)(1+r/r_s)^2} \, ,
\end{align}
where $\rho_0=0.33\ \rm GeV/cm^3$, $r_s=20$ kpc and $r$ is the distance from a galactic source point at location $(s,l,b)$ to Earth: 
\begin{align}
r(s,l,b) \ = \ \sqrt{s^2 + R_{\oplus}^2-2\,sR_{\oplus}{\rm cos}\:(b){\rm cos}\:(l)} \, , 
\end{align}
with $R_{\oplus}=8.5$ kpc  being  the  distance from the Sun to the galactic center. Similarly, the extragalactic contribution, averaged over all directions, is given by 
\begin{equation}
\Phi_{\rm EG} \ = \ \frac{\Omega_{\rm DM}\cdot \rho_c}{4\pi\cdot  M_{\rm DM}\cdot\tau_{\rm DM}}\int_{0}^{\infty} \frac{dz}{H(z)}\frac{dN((1+z)E_\nu)}{dE_{\nu}} \, ,
\end{equation}
where $\Omega_{\rm DM} = 0.27$, $\rho_c=5.5\times10^{-6}\ \rm GeV/cm^3$ and $H(z)=H_0 \sqrt{\Omega_{\Lambda}+\Omega_{\rm M} (1+z)^3}$ with $H_0=67\ \rm km \: Mpc^{-1} s^{-1}$, $\Omega_{\Lambda}=0.68$ and $\Omega_{\rm M}=0.32$~\cite{Patrignani:2016xqp}. 

For both galactic and extragalactic fluxes, we have $dN(E_\nu)/dE_{\nu}$ representing the neutrino spectrum per decay at source point, which in principle is a function of ($M_{\rm DM},\tau_{\rm DM}$). We calculate the neutrino spectrum from DM decay in our model by implementing it in MADGRAPH2.6~\cite{Alwall:2014hca} and passing the generated events through PYTHIA8.2~\cite{Sjostrand:2007gs} for parton showering, with the electroweak radiation effects taken into consideration.  We repeat this procedure by scanning over different ($M_{\rm DM},\tau_{\rm DM}$) values to generate the parameter space in the ($M_{\rm DM},\tau_{\rm DM}$) plane which could explain the IceCube data. 


\subsection{Fit Results}
We follow the same procedure as outlined in Section~\ref{sec:reconstruct} and obtain the test statistics using Eq.~\eqref{eq:TS}, but now with four free parameters:  $\theta=\{M_{\rm DM},\tau_{\rm DM},\Phi_0,\gamma_0\}$ and with  
\begin{equation}\label{eq:dmevent}
N^{\rm HESE}_{{\rm tot},i} \ = \ N^{\rm HESE}_{{\rm DM},i}+N^{\rm HESE}_{{\rm astro},i}+N^{\rm HESE}_{{\rm atm},i} \, ,
\end{equation}  
and similarly for $N^{\rm TG}_{{\rm tot},i}$. 
Our bestfit results, along with the corresponding TS/dof, are given in Table~\ref{table:2} for both $(1:1:1)$ and $(4:7:7)$ flavor compositions for the astrophysical neutrino flux, while for the neutrino flux coming from DM, we have assumed a simple $(1:1:1)$ flavor ratio. Just based on the TS/dof, both $(1:1:1)$ and $(4:7:7)$ cases for the astrophysical neutrino flux provide a good fit to the combined HESE and TG data. The fit is better than the purely two-component astrophysical flux, cf. Table~\ref{tab:1}.  

\begin{table}[t!]
	\centering
	\begin{tabular}{c c| c c c c | c } 
		\hline\hline
		 DM (1st comp.) & astro (2nd comp.) & $\Phi_{0}$ & $\gamma_0$ & $M_{\rm DM}~(\rm TeV)$ & $\tau_{\rm DM}(10^{28}~{\rm s})$ & TS/dof \\ [0.5ex] 
		\hline
		$(1:1:1)$ & $(1:1:1)$ & 1.62 & 2.00 & 316.23 & 6.31 & 1.38 \\ 
		$(1:1:1)$ & $(4:7:7)$ & 1.39 & 1.97 & 316.23 & 6.31 & 1.37 \\[1ex] 
		\hline\hline
	\end{tabular}
	\caption{Bestfit results for the DM+astrophysical two-component neutrino flux, cf.~Eq.~\eqref{eq:dmastro}. Here $\Phi_{0}$ is in units of $10^{-18}/({\rm GeV\: sr\: s\: cm}^2)$.}
	\label{table:2}
\end{table}
\begin{figure}[t!]
	\centering
	\begin{subfigure}[b]{0.40\textwidth}
		\centering
		\includegraphics[width=\textwidth]{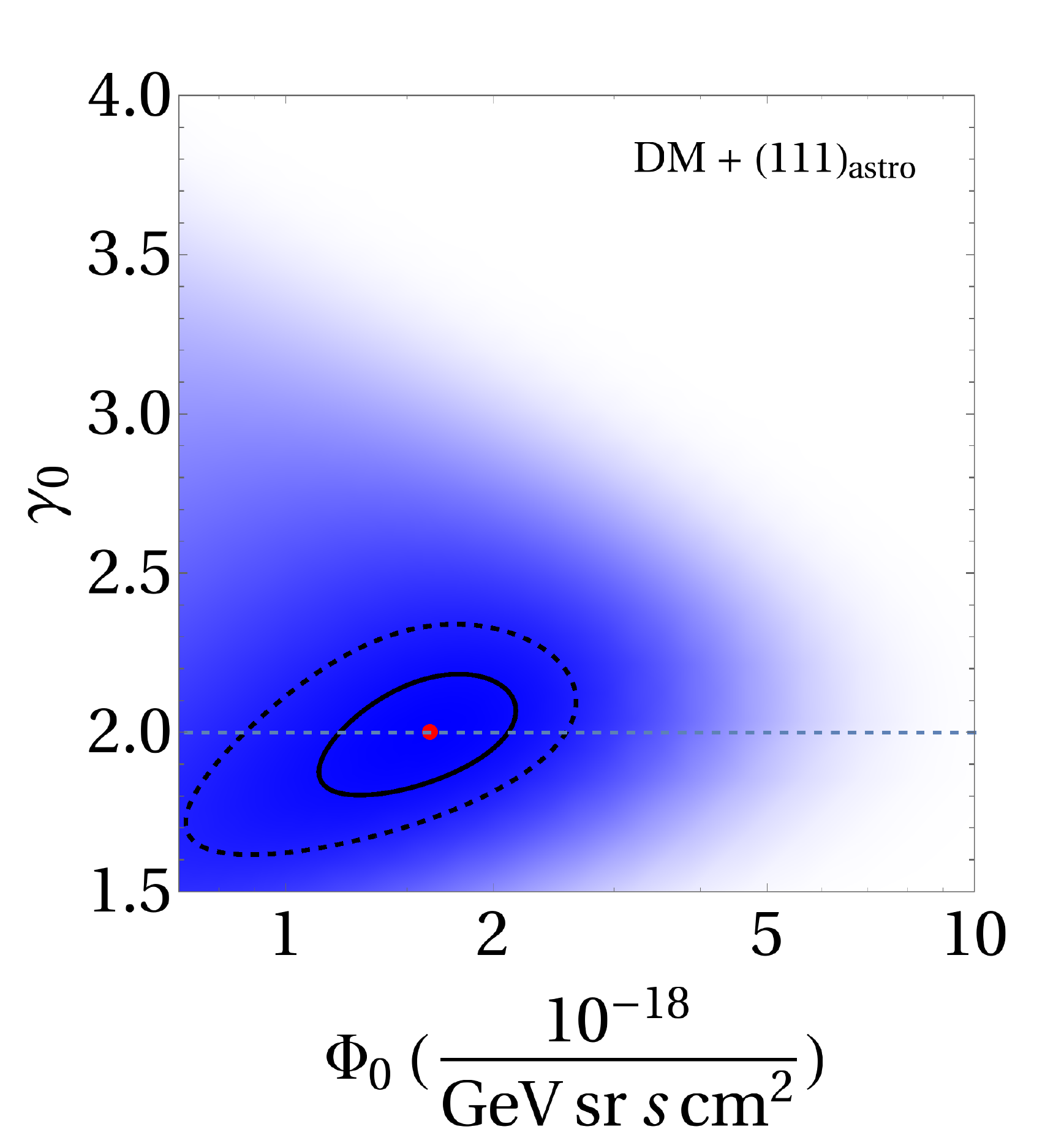}
		\caption[Network2]%
		{{\small $(\Phi_{0},\gamma_0)$ map for DM+(1:1:1)$_{\rm astro}$.}}    
	\end{subfigure}
	\quad
	\begin{subfigure}[b]{0.40\textwidth}  
		\centering 
		\includegraphics[width=\textwidth]{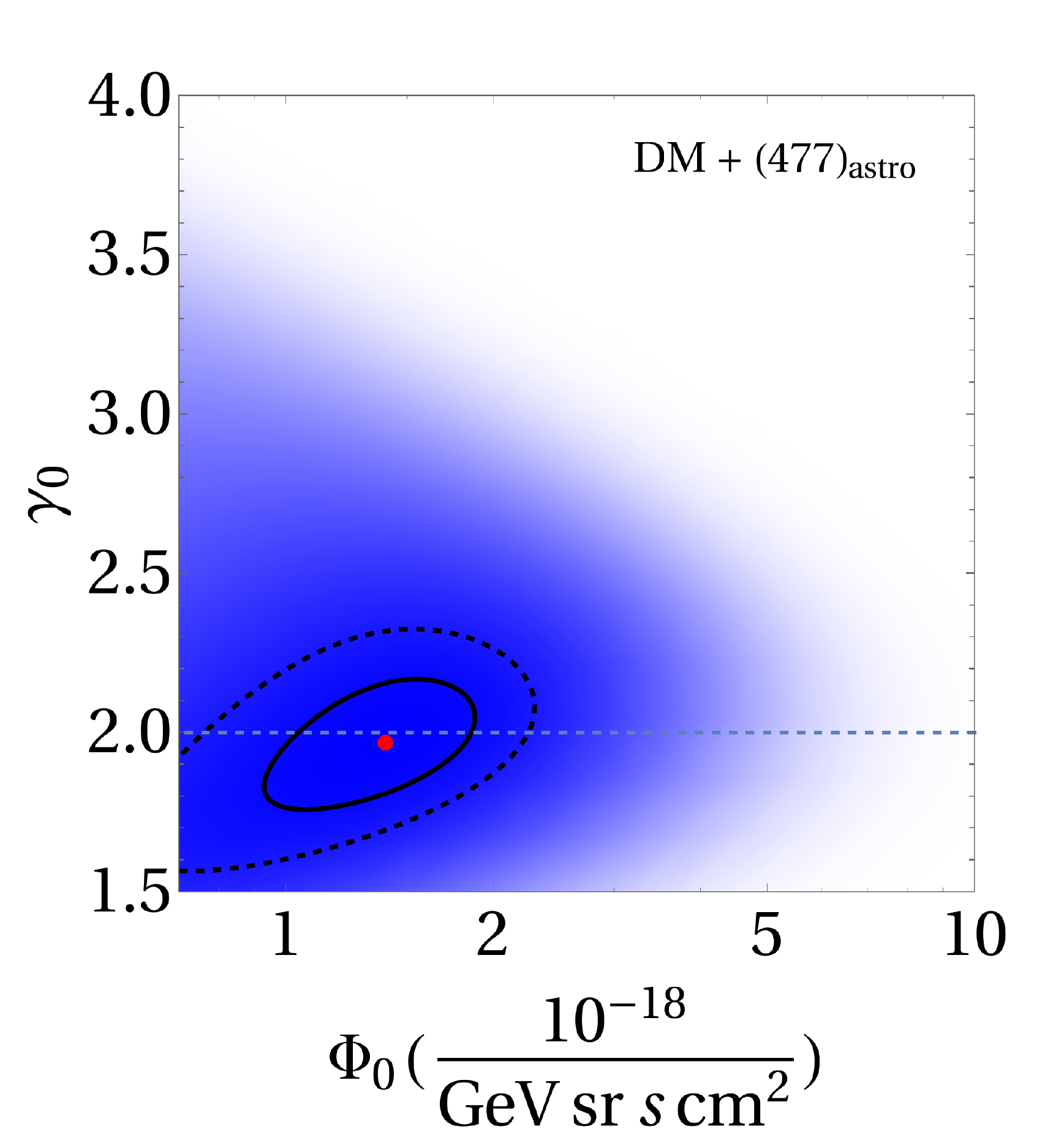}
		\caption[]%
		{{\small $(\Phi_{0},\gamma_0)$ map for DM+(4:7:7)$_{\rm astro}$.}}    
	\end{subfigure}
	\vskip\baselineskip
	\begin{subfigure}[b]{0.40\textwidth}   
		\centering 
		\includegraphics[width=\textwidth]{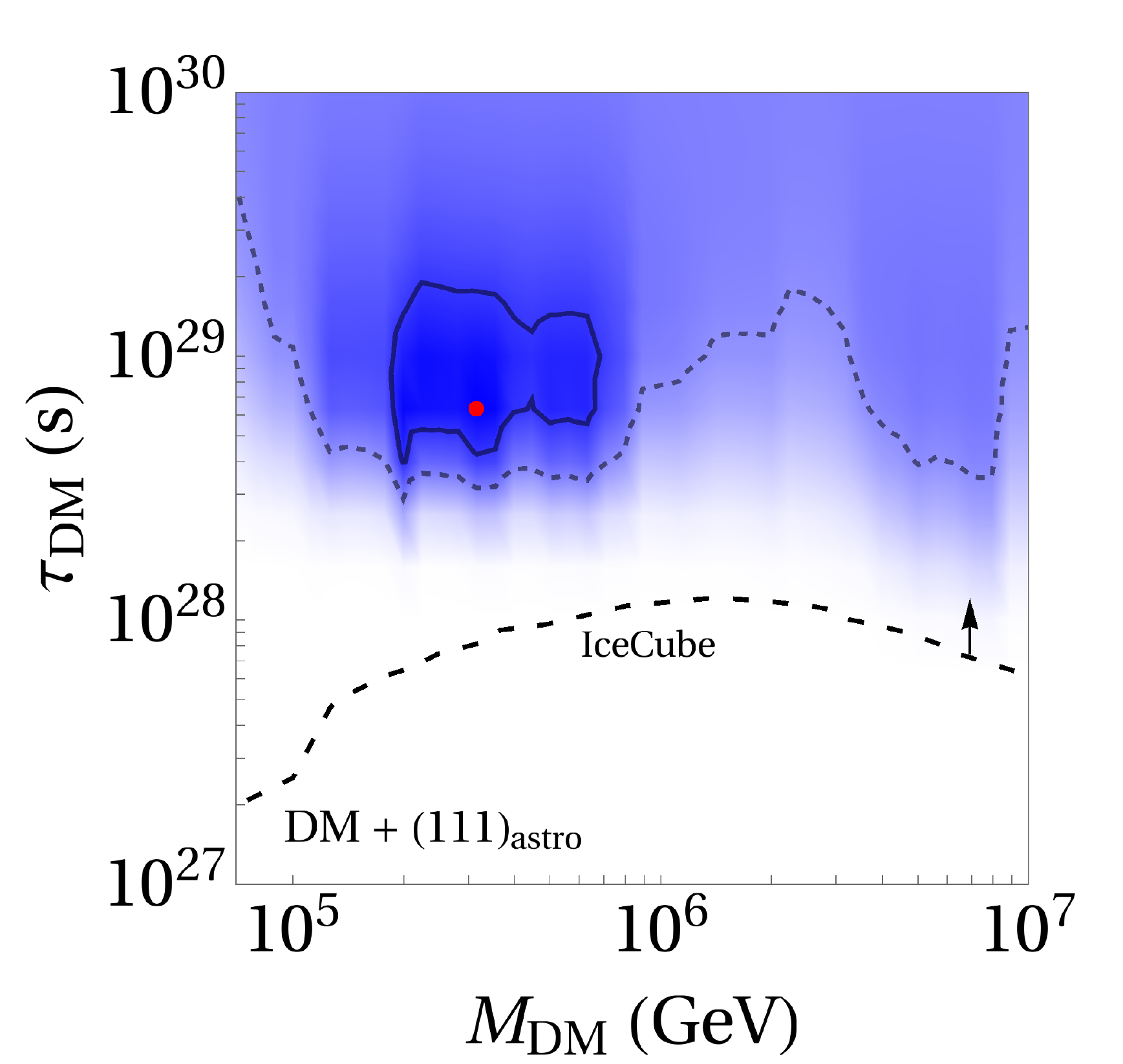}
		\caption[]%
		{{\small Mass ($M_{\rm DM}$) and decay time ($\tau_{\rm DM}$) map for DM+(1:1:1)$_{\rm astro}$.}}    
	\end{subfigure}
	\quad
	\begin{subfigure}[b]{0.40\textwidth}   
		\centering
		\includegraphics[width=\textwidth]{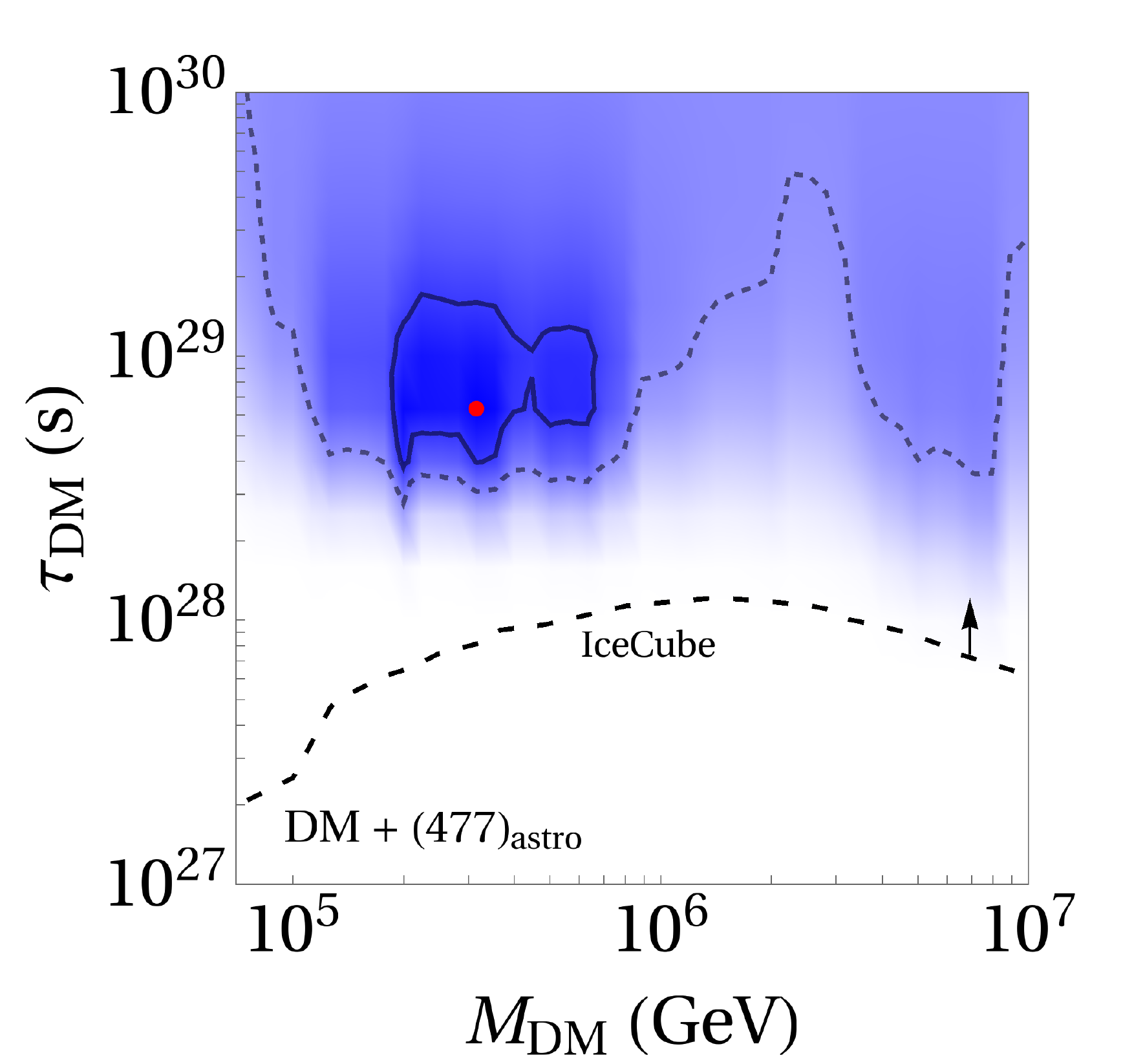}
		\caption[]%
		{{\small Mass ($M_{\rm DM}$) and decay time ($\tau_{\rm DM}$) map for DM+(4:7:7)$_{\rm astro}$.}}    
	\end{subfigure}
	\caption[]
	{\small Preferred regions of parameter space in the $(\Phi_{{0}},\gamma)$ [top panels] and $(M_{\rm DM},\tau_{\rm DM})$ [bottom panels] planes for the DM+astrophysical two-component neutrino flux, cf. Eq.~\eqref{eq:dmastro}, with $(1:1:1)$ [left panels] and $(4:7:7)$ [right panels] astrophysical neutrino flavor compositions. The solid and thin dashed contours show the $2\sigma$ and 3$\sigma$ preferred ranges respectively, while the red dot in each map shows the bestfit value as given in Table~\ref{table:2}. The thick black dashed curve in the bottom panels is the 90\% CL lower limit on the DM lifetime derived by the IceCube collaboration~\cite{Aartsen:2018mxl}.} 
	\label{fig:1compDMmap}
\end{figure}

In Figure~\ref{fig:1compDMmap}, we show the preferred regions of parameter space in the $(\Phi_{{0}},\gamma)$ [top panels] and $(M_{\rm DM},\tau_{\rm DM})$ [bottom panels] planes for the DM+astrophysical two-component neutrino flux, cf. Eq.~\eqref{eq:dmastro}, with $(1:1:1)$ [left panels] and $(4:7:7)$ [right panels] astrophysical neutrino flavor compositions. The solid and thin dashed contours show the $2\sigma$ and 3$\sigma$ preferred ranges respectively, while the red dot in each map shows the bestfit value as given in Table~\ref{table:2}. The thick black dashed curve in the bottom panels is the 90\% CL lower limit on the DM lifetime recently derived by the IceCube collaboration~\cite{Aartsen:2018mxl} using 6 years of IceCube data focusing on muon neutrino
`track' events from the Northern Hemisphere and two years of `cascade' events from the full
sky. The IceCube limit significantly improves upon the previous best limits from gamma-rays~\cite{Ackermann:2012rg, Cohen:2016uyg}, neutrinos~\cite{PalomaresRuiz:2007ry, Abbasi:2011eq}, cosmic rays~\cite{Gondolo:1991rn} and cosmic microwave background (CMB) radiation~\cite{Cline:2013fm}. Our $3\sigma$ preferred region to fit both HESE and TG data is consistent with the recent IceCube limit. 
 
\begin{figure}[t!]
	\centering
	\begin{subfigure}[b]{0.48\textwidth}
		\centering
		\includegraphics[width=\textwidth]{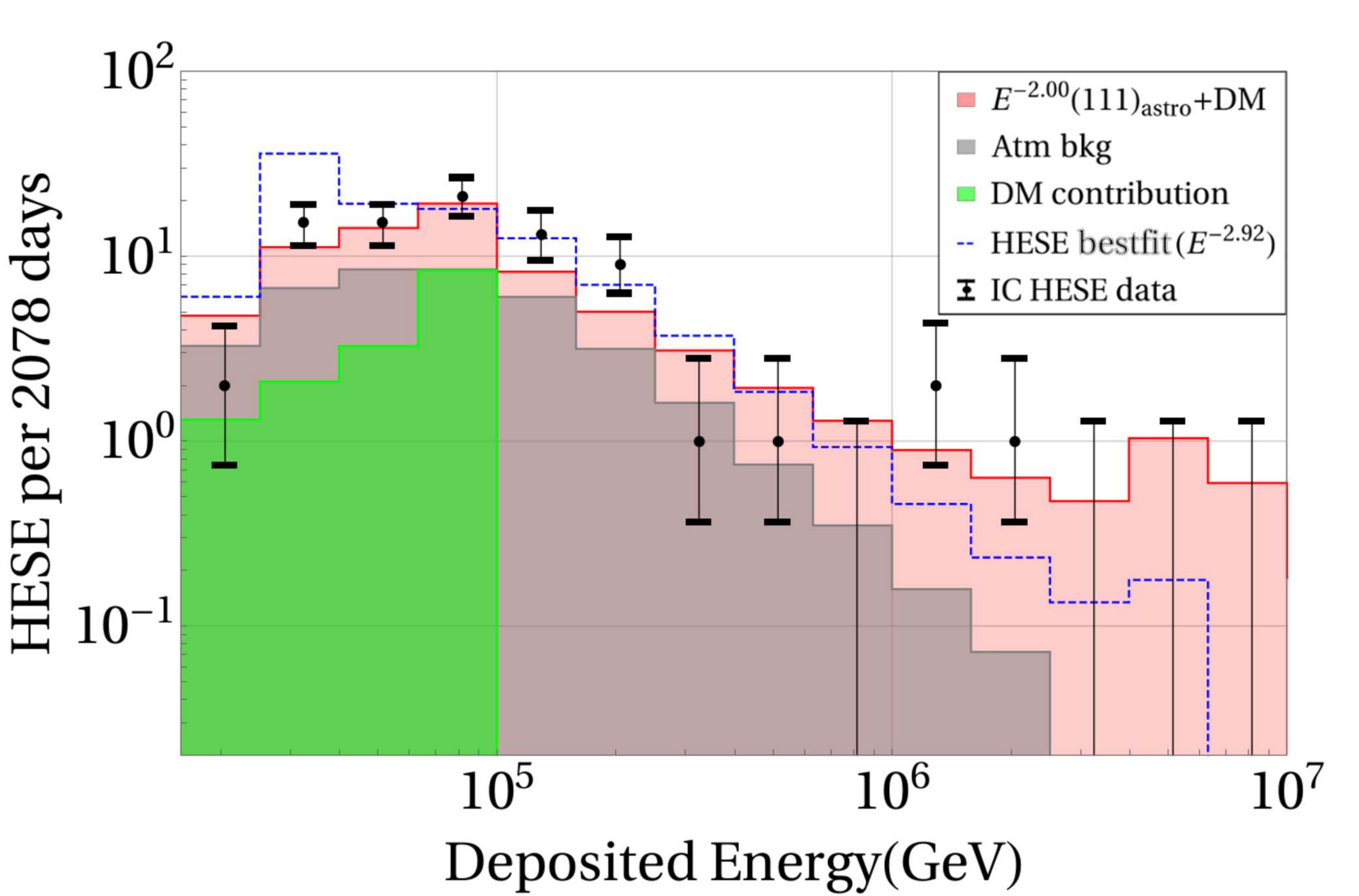}
		\caption[]%
		{{\small Combined bestfit predictions for HESE with $(1:1:1)_{\rm DM}+(1:1:1)_{\rm astro}$ and comparison with the one-component fit.}}    
		\label{fig:1111111compDMEventHESE}
	\end{subfigure}
	\quad
	\begin{subfigure}[b]{0.48\textwidth}  
		\centering 
		\includegraphics[width=\textwidth]{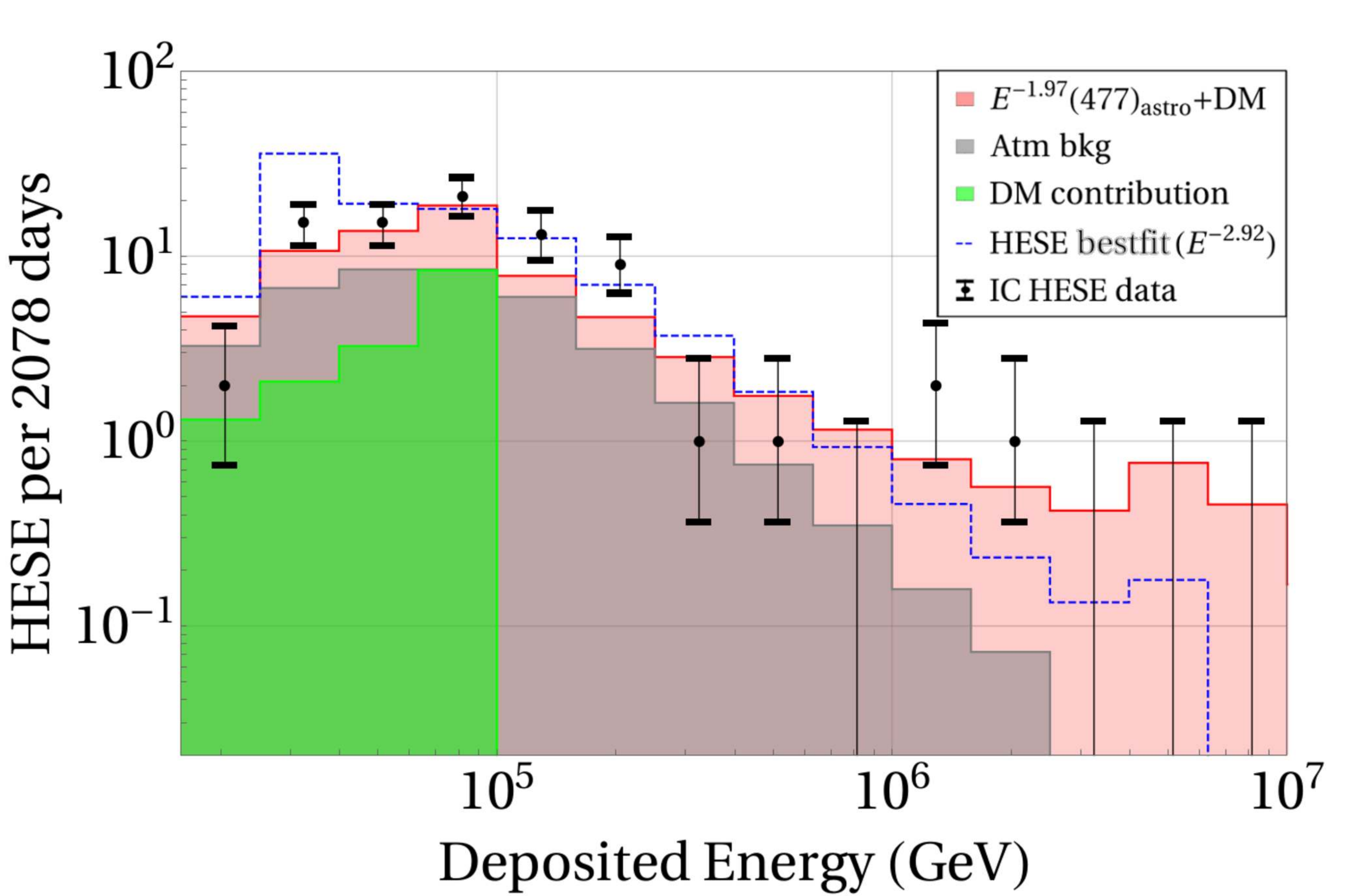}
		\caption[]%
		{{\small Combined bestfit predictions for HESE with $(1:1:1)_{\rm DM}+(4:7:7)_{\rm astro}$ and comparison with the one-component fit.}}    
		\label{fig:1114771compDMEventHESE}
	\end{subfigure}
	\vskip\baselineskip
	\begin{subfigure}[b]{0.48\textwidth}   
		\centering 
		\includegraphics[width=\textwidth]{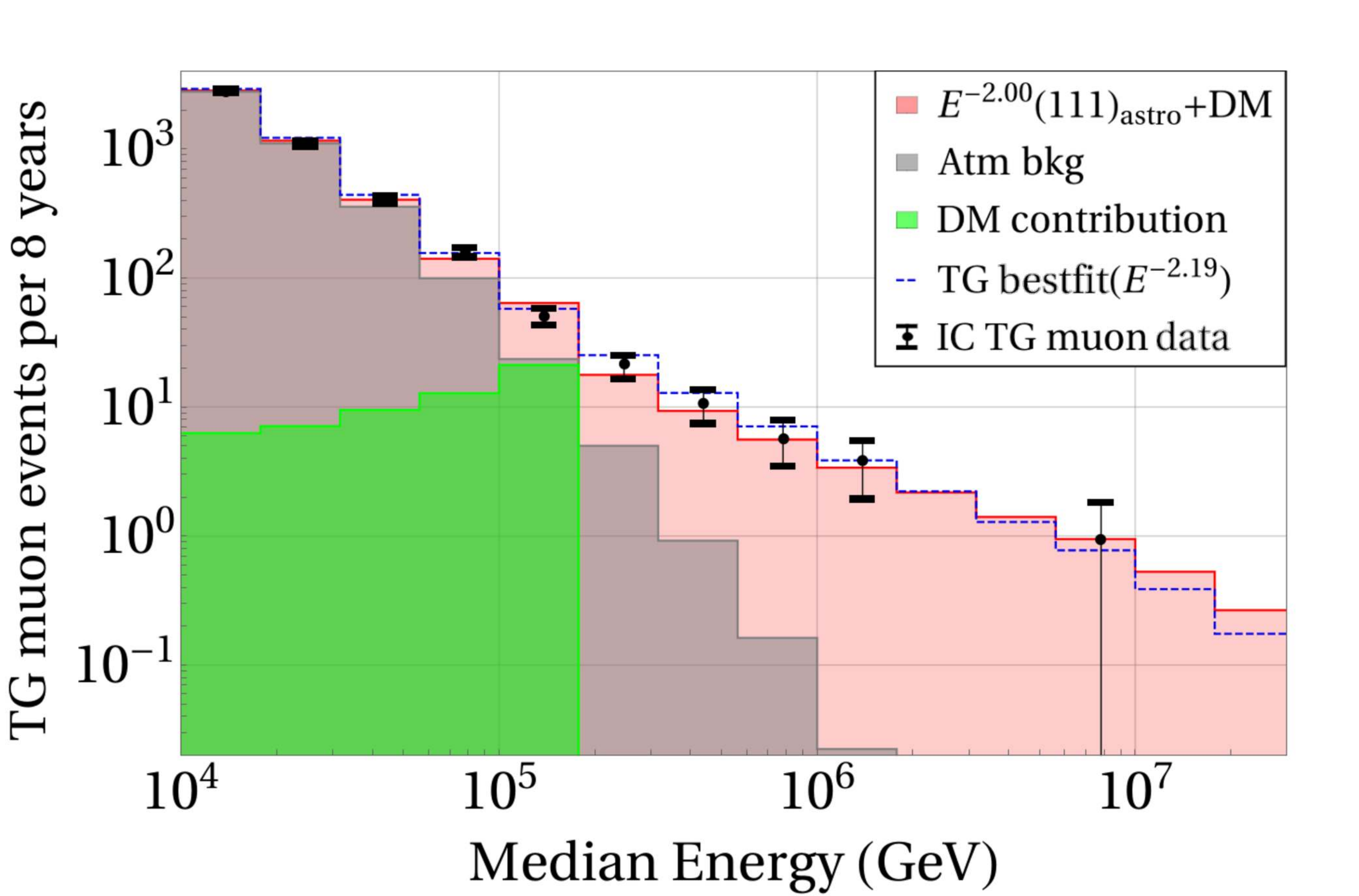}
		\caption[]%
		{{\small Combined bestfit predictions for TG muons with $(1:1:1)_{\rm DM}+(1:1:1)_{\rm astro}$ and comparison with the one-component fit.}}    
		\label{fig:1111111compDMEventTG}
	\end{subfigure}
	\quad
	\begin{subfigure}[b]{0.48\textwidth}   
		\centering 
		\includegraphics[width=\textwidth]{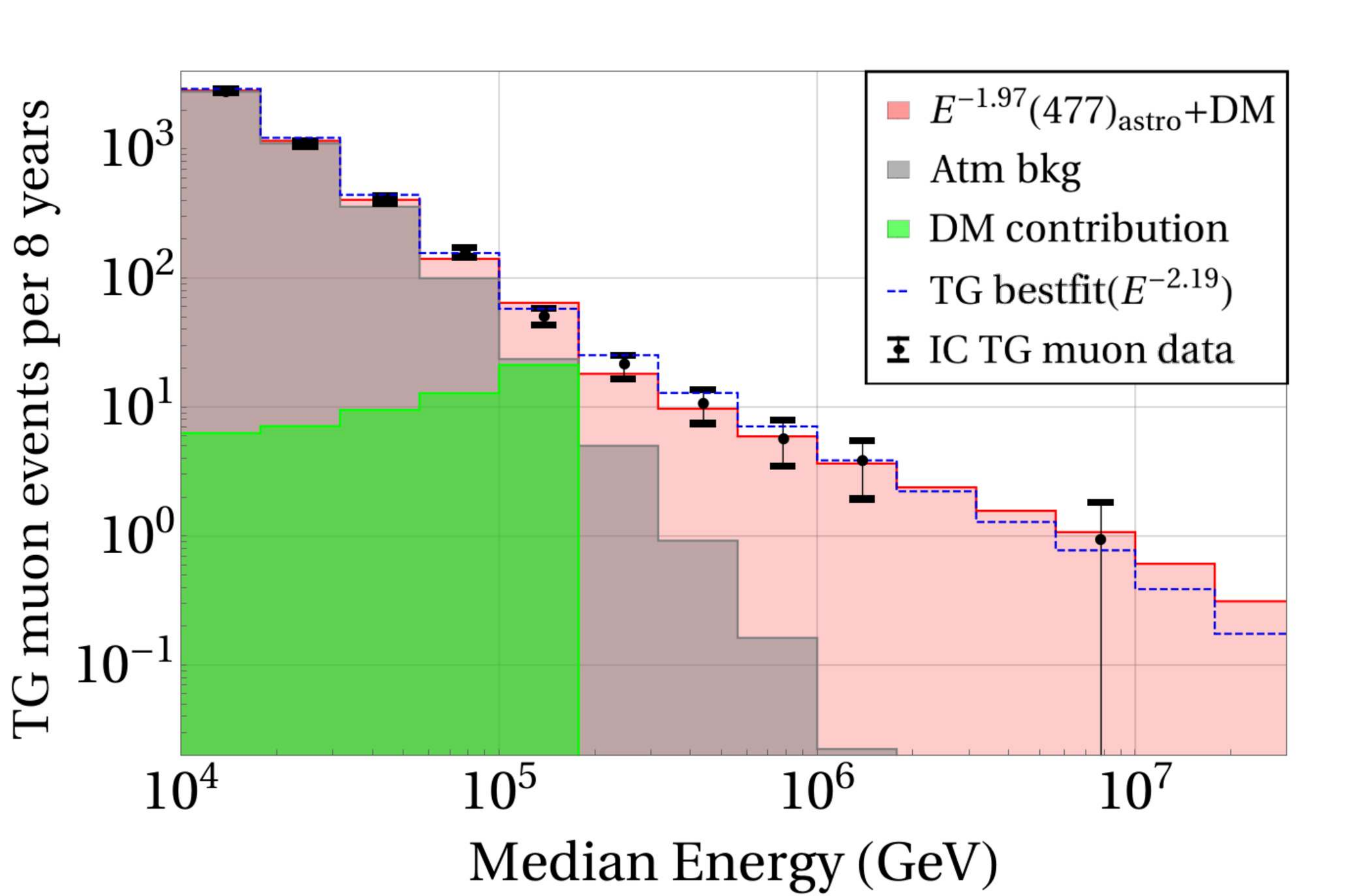}
		\caption[]%
		{{\small Combined bestfit predictions for TG muons with $(1:1:1)_{\rm DM}+(4:7:7)_{\rm astro}$ and comparison with the one-component fit.}}    
		\label{fig:1114771compDMEventTG}
	\end{subfigure}
	\caption[]
	{\small Event spectrum for the DM+astro two-component bestfit reconstruction of HESE and TG data samples. The grey shaded part is the contribution due to atmospheric background and the pink shaded part is the total contribution (two-compopnet+ bkg), with the green shaded part the individual contribution from the DM component. The dashed blue curve in (a) and (b) is the one-component HESE-alone bestfit [cf.~Eq.~\eqref{eq:1compfit}], while the dashed blue curve in (c) and (d)  is the one-component TG-alone bestfit [cf.~Eq.~\eqref{eq:1compfit2}]. The IceCube data points for 6-year HESE and 8-year TG samples are also shown.} 
	\label{fig:1compDMEvents}
\end{figure}
In Figure~\ref{fig:1compDMEvents}, we present the DM+astro bestfit reconstructed event spectrum for both HESE and TG data samples. The atmospheric background, one-component fit and the IceCube data are the same as in Figure~\ref{fig:2compEvents}. The main difference is that now the DM decay contribution (green shaded) serves as the low-energy part of the two-component flux, with a natural energy cut-off at $M_{\rm DM}/2$ due to kinematic reasons. This also provides a good explanation of the apparent excess in the vicinity of 100 TeV in the HESE data. The high-energy component could be either $(1:1:1)$ or $(4:7:7)$ astrophysical neutrino flux. The combined flux provides a very good fit to both HESE and TG data simultaneously.  Moreover, the bestfit spectral index for the astrophysical component miraculously turns out to be $\gamma=2$ purely from the data-fitting, which is the same value as expected theoretically in the Fermi shock model.  

From the event spectrum of Figure~\ref{fig:2compEvents}, we can also understand certain features of the bestfit given in Table~\ref{table:2}. For instance, the DM parameters remain unchanged for the two cases, which is understandable since in the 100 TeV bump area, the DM flux contributes the most. It is the location and peak height within this short range of energy that decides $M_{\rm DM}$ and $\tau_{\rm DM}$, independent of the high-energy component. Similarly, the $(4:7:7)$ case has a smaller normalization factor because it has already got a larger portion of $\nu_\mu$ at high energy to fit the TG data, thus it no longer needs a flux as large as the $(1:1:1)$ case.

\section{Multi-messenger Constraints from Gamma Ray Flux}\label{sec:gamma}

As discussed in Section~\ref{sec:astro}, the astrophysical neutrinos are produced via hadronic interactions of the cosmic rays, such as $pp$, $pn$ or $p\gamma$, which lead to the production of mesons, mainly $\pi^\pm$ and $\pi^0$. While the charged pion decays give rise to the neutrinos via weak interactions, the neutral pions promptly decay to photons via electromagnetic interaction: $\pi^0\to 2\gamma$. The energy of the resultant $\gamma$-rays is, on average, $E_\gamma=2E_\nu$, taking into consideration the energy correlations: $E_{\gamma}=E_{\pi^0}/2 ,\ E_\nu=E_{\pi^\pm}/4$ and $E_{\pi^0}\simeq E_{\pi^\pm}$. This leads to a relation between the photon and neutrino fluxes as well~\cite{Waxman:1997ti, Murase:2013rfa, Murase:2015xka}. For the canonical pion and muon decay source, this can be approximated by the following~\cite{Meszaros:2017fcs}:  
\begin{equation}
\label{eq:fluxeq}
E^2_\gamma\Phi_\gamma \ \simeq \ \frac{4}{K} E^2_{\nu}\frac{\Phi_{(\nu+\bar{\nu})_{\rm tot}}}{3}\bigg|_{E_\nu=0.5{E_\gamma}}
\end{equation}
where $\Phi_{(\nu+\bar{\nu})_{\rm tot}}$ stands for the total $\nu+\bar{\nu}$ flux (summed over all flavors), and $K$ is the ratio of the charged to neutral pions, which is $K\simeq 1$ for the $p\gamma$ case and $K\simeq 2$ for the $pp$ case. For the muon-damped pion decay source, one should use Eq.~\eqref{eq:fluxeq} without the factor of $1/3$, because for the same neutrino flux, the photon flux is larger by a factor of 3 in this case. 




As for the two-component case, we no longer have the simple  correlation between the photon and neutrino fluxes as in Eq.~\eqref{eq:fluxeq}, since it will now be affected by the energy-dependent ratio of fluxes between the two components ($\Phi_1(E_\nu)/\Phi_2(E_\nu)$). However, we can calculate all the possible combinations of the individual components with different $K$ values and choose the maximum and minimum values of the fluxes as a conservative estimate, because the energy-dependent correlation will lie somewhere between these two extrema. Similarly, for the DM case, we calculate the photon flux corresponding to a given neutrino flux numerically using PYTHIA and add it to the astrophysical contribution. Finally, we integrate both sides of Eq.~\eqref{eq:fluxeq} to obtain the integrated photon flux for a given bestfit neutrino flux and check its compatibility with the existing upper limits for the multi-messenger constraints.



It is also important to consider attenuations of the photon flux for both galactic and extragalactic sources, unlike the neutrino flux which is hardly attenuated due to the weakness of neutrino interactions. Following the arguments in Ref.~\cite{Esmaili:2015xpa}, we assume that the photons from extragalactic sources  are fully absorbed due to the $e^+ e^-$ pair production with CMB and EBL in the propagation process. This means that only the galactic component of the photon flux is relevant for us. Given that the galactic contribution to the IceCube neutrino flux is at most 14\%~\cite{Aartsen:2017ujz}, we use 14\% of the total galactic integrated gamma-ray flux prediction from Eq.~\eqref{eq:fluxeq} to derive the multi-messenger constraints.  The gamma-ray referred to here is of course the diffuse isotropic component. All the high-energy gamma-ray beams from specific nearby astrophysical sources are not of concern, since the neutrino spectrum detected by IceCube shows an isotropic feature, not spatially correlated with any resolved point sources. Moreover, even the galactic gamma-ray flux will suffer from attenuation for $E_\gamma \gtrsim 10$ TeV. Based on the dynamics with CMB photons, the optical depth as a  function of $E_\gamma$ is used to estimate these attenuation effects~\cite{Esmaili:2015xpa}.

\begin{figure}[t!]
	\centering
	\begin{subfigure}[b]{1.0\textwidth}
		\centering
		\includegraphics[width=0.6\textwidth]{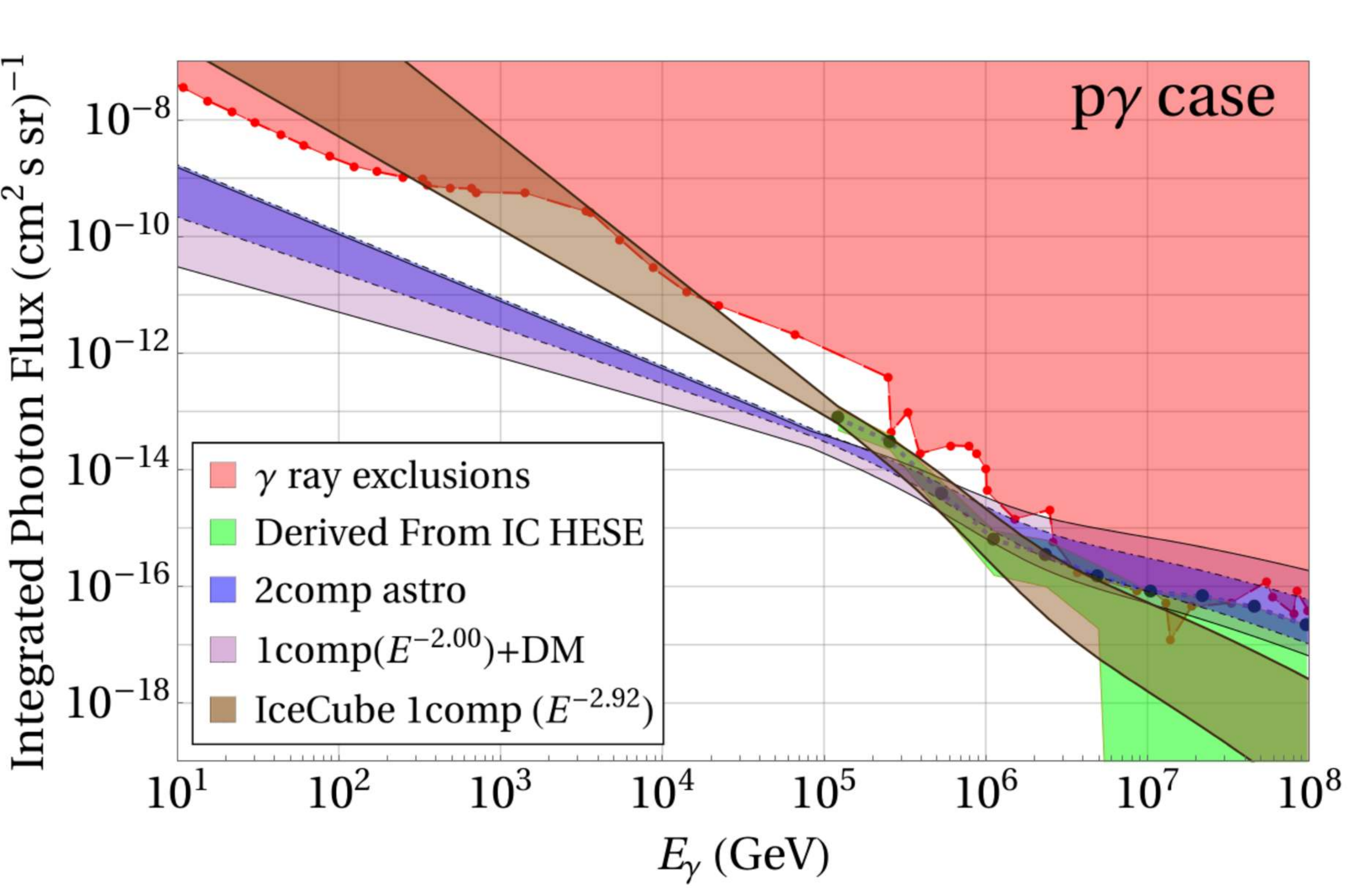}
		\caption[Network2]%
		{{\small $K=1$ case}}    
	\end{subfigure}
	\quad
	\begin{subfigure}[b]{1.0\textwidth}  
		\centering 
		\includegraphics[width=0.6\textwidth]{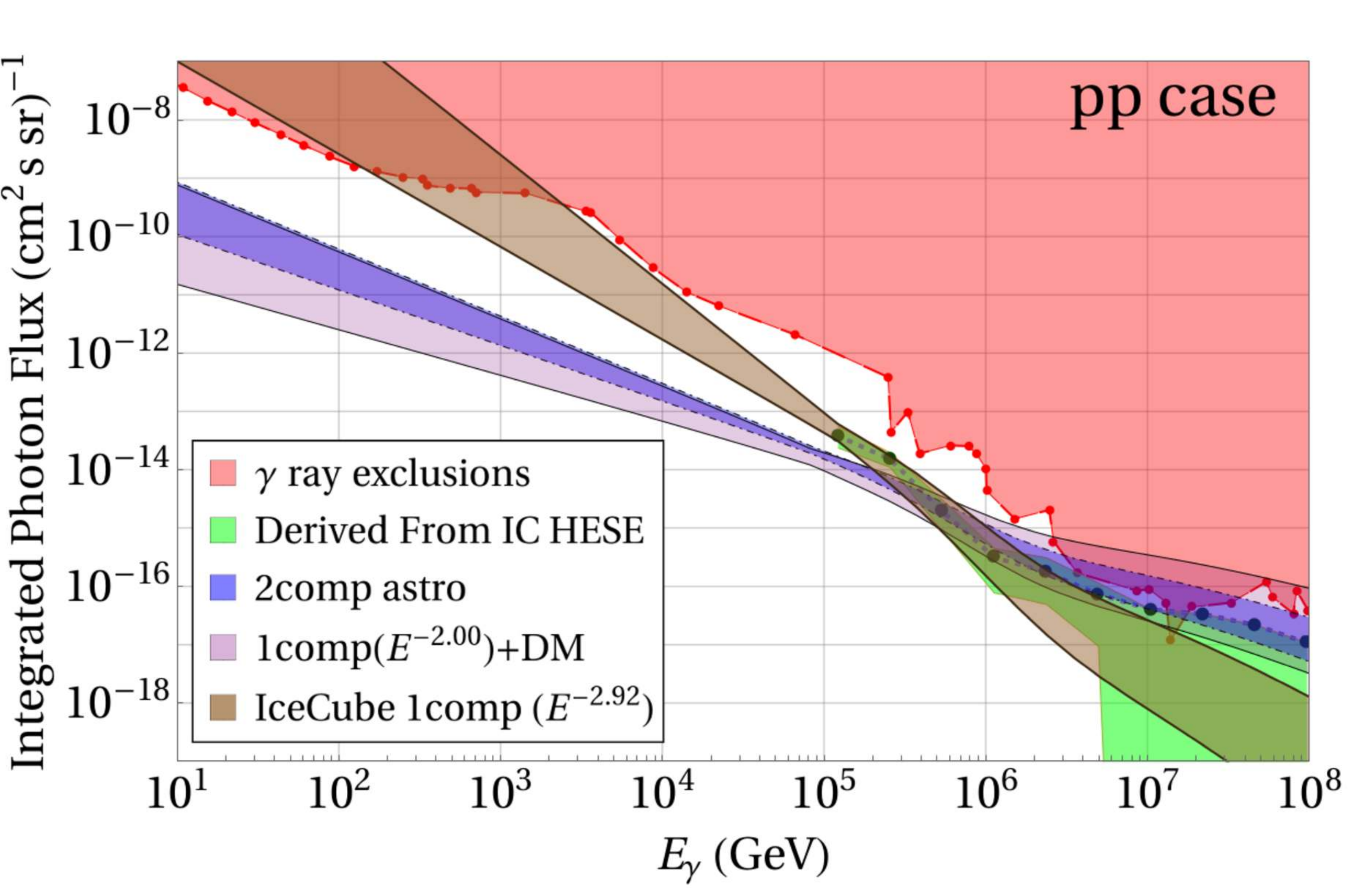}
		\caption[]%
		{{\small $K=2$ case }}    
	\end{subfigure}
	\caption[]
	{\small The integrated photon flux predictions corresponding to the $2\sigma$ range around the bestfit one-component (brown), two-component astrophysical (blue) and DM+astrophysical (magenta) neutrino fluxes for the $p\gamma$ (upper panel) and $pp$ (lower panel) cases. Here all flavor compositions have been taken as $(1:1:1)$ for illustration. The twist in the bestfit regions around PeV is caused by the galactic attenuation effect. The red-shaded region is the 90\% CL exclusion from the combination of the existing gamma-ray constraints, with the experimental data points shown by red dots.} 
	\label{fig:photonconst}
\end{figure}

Our multi-messenger constraint results are shown in Figure~\ref{fig:photonconst}. The top panel is for $K=1$ ($p\gamma$) and the bottom panel is for $K=2$ ($pp$). In each panel, we show the current 90\% CL upper limits on the integrated photon flux over the whole energy range from 10 GeV to $10^8$ GeV, obtained from a combination of various experiments, such as CASA-MIA~\cite{Chantell:1997gs}, MILARGO~\cite{Abdo:2008if}, Fermi-LAT~\cite{Ackermann:2014usa},  GRAPES~\cite{Gupta:2014nya}, 
KASCADE~\cite{Feng:2015dye,Apel:2017ocm}, ARGO~\cite{DiGirolamo:2016kur} and HAWC~\cite{Pretz:2015wma}. We also adopted a combined best fit~\cite{Lipari:2018gzn} in the 1-100 TeV region for the upper bound on photon flux based on the diffuse electron flux limit from AMS-02~\cite{Aguilar:2016kjl}, DAMPE~\cite{Ambrosi:2017wek}, Fermi-LAT~\cite{Abdollahi:2017nat}, MAGIC~\cite{BorlaTridon:2011dk}, HESS~\cite{Abdalla:2017brm},  and VERITAS~\cite{Staszak:2015kza}, since air showers from electrons, positrons and photons behave the same. The light brown region corresponds to the $2\sigma$ region for the IceCube HESE-only bestfit for the one-component flux. It is clear that the HESE-only bestfit is in severe conflict with the gamma-ray constraint, especially in the low-energy region. This is mainly due to the softer neutrino spectrum, cf.~Eq.~\eqref{eq:1compfit}, which predicts more neutrinos in the lower energy range, thus leading to an increased photon flux as well. Note that the inverse Compton photon flux was not included in our analysis, which contributes to the total photon flux mainly below $\sim 100$ GeV. Including this effect will further aggravate the tension between the IceCube bestfit and the photon constraint. 

Our $2\sigma$ range for the astrophysical two-component bestfit is shown in Figure~\ref{fig:photonconst} by the blue region. This is largely consistent with the gamma-ray limits, except that there is a mild tension around 10 PeV. For the $p\gamma$ case, the tension starts around 1 PeV, because in this case, we expect more photons compared to the $pp$ case for the same neutrino flux. On the other hand, for the DM+astrophysical two-component case (magenta region), the tension is almost gone. This is mainly due to the harder neutrino spectrum, which predicts less neutrino events in the low-energy range, and also because of the neutrino-dominated DM decay which produces less high-energy photons. Although it leads to a slightly higher number of events at high energy, this effect is minimized by a combination of the lower flux and attenuation effects. For the $(4:7:7)$ astrophysical case, the results are similar, so not shown here. In the future, with more data from multi-messenger probes, these two-component scenarios could be decisively tested and distinguishable from each other.

Before closing, a few comments are in order: 
\begin{itemize}
\item [(i)] For the photon constraints, we only show the energy range till $10^8$ GeV. This is because the astrophysical neutrinos and their photon counterparts have roughly $\sim 1-5\%$ of the typical CR energy. Thus for the CR having energy right before the GZK cutoff of $10^{11}$ GeV~\cite{Greisen:1966jv, Zatsepin:1966jv}, the corresponding $\nu$ and $\gamma$ fluxes should have the maximum energy around $10^8$ GeV. 

\item [(ii)] For the DM decay, apart from the $\gamma$ flux comparison, we also did an antiproton flux comparison with the recent data from AMS-02~\cite{Aguilar:2016kjl}. The specific DM decay channels available in our model here do not give rise to significant antiproton flux and are well within the AMS-02 constraint. It might be interesting to examine other possible decaying DM models with different decay final states (see e.g.~Refs.~\cite{Murase:2015gea, Kalashev:2016cre, Cohen:2016uyg, Aartsen:2018mxl}) in light of the antiproton constraints, in addition to the photon constraints. 

\item [(iii)] It was recently shown in Ref.~\cite{Murase:2016gly} that a UHE neutrino flux with $\gamma=2$ is simultaneously consistent with the multi-messenger neutrino, gamma-ray and UHECR constraints; see also Ref.~\cite{Fang:2017zjf} for a specific astrophysical example model. Our bestfit two-component solutions corroborate this finding. 
\end{itemize}


\section{Conclusion} \label{sec:conc}
In this paper, we have explored the possibility of using a two-component neutrino flux model to simultaneously explain the IceCube HESE and throughgoing muon events above 10 TeV. We have considered two different types of two-component flux: (i) purely astrophysical, and (ii) one-component astrophysical plus a decaying dark matter. In both cases, we also consider two different flavor compositions on Earth for the astrophysical neutrinos, namely $(1:1:1)$ and $(4:7:7)$. In each case, we perform a likelihood analysis with the combined HESE and TG data samples to determine the bestfit spectral indices, flux normalization, cut-off energy for the first component, and in the DM decay case, the mass and lifetime of the DM. For the two-component astrophysical case, we find that the data prefers the high energy component being $(4:7:7)$, mainly due to the absence of Glashow events. On the other hand, the two-component astrophysical case does not give a good fit to the HESE data, especially in the 100 TeV region. We show that this issue can be addressed by replacing the low-energy astrophysical component with a decaying DM component. We obtain the best fit values of $M_{\rm DM}\:=\:315^{+335}_{-125}~\rm TeV$ and $\tau_{\rm DM}\:=\:6.3^{+12.7}_{-2.3}\times 10^{28}~\rm s$ for the DM mass and lifetime respectively. Moreover, in this case, the spectral index for the astrophysical component is consistent with the theoretical prediction of $\gamma=2$. 

We have also checked the compatibility of the two-component fluxes with the multi-messenger constraints from integrated diffuse photon flux observations. Using the robust correlation between astrophysical neutrino and photon fluxes, we show that the HESE-only IceCube bestfit is clearly ruled out by the gamma-ray constraints. The bestfit two-component astrophysical flux is still safe, but on the verge of being excluded. The DM+astrophysical flux seems to be the best solution for the current data, while being consistent with the photon constraints.

\section{Acknowledgments} 

We gratefully acknowledge discussions with James Buckley, Ramanath Cowsik, Raj Gandhi, Ranjan Laha, Danny Marfatia, Kohta Murase, and Yongchao Zhang at various stages of this work. BD would also like to thank the organizers of the WHEPP XV at IISER Bhopal, where part of this work was done, for the local hospitality.


\end{document}